\documentclass[12pt]{iopart}

\usepackage{graphicx,cite,amssymb}

\usepackage{color}

\newcommand{\ii}{\mathrm{i}}

\newcommand{\dt}{\ensuremath{{\delta t}}}

\let \Re \relax
\newcommand{\Re}{\mathrm{Re}\,}

\newcommand{\eqref}[1]{(\ref{#1})}

\begin{document}

\title{
Numerical time propagation of quantum systems in radiation fields
}

\author{A Alvermann$^1$\footnote{Present address: Institut f\"ur Physik, Ernst-Moritz-Arndt-Universit\"at, 17487 Greifswald, Germany}, H Fehske$^2$ and P B Littlewood$^3$}

\address{$^1$ Theory of Condensed Matter, Cavendish Laboratory, Cambridge CB3 0HE, United Kingdom}

\address{$^2$ Institut f\"ur Physik, Ernst-Moritz-Arndt-Universit\"at, 17487 Greifswald, Germany}

\address{$^3$ Argonne National Laboratory, Argonne, IL60439, USA}

\ead{alvermann@physik.uni-greifswald.de}

\begin{abstract}

Atoms, molecules or excitonic quasiparticles,
for which excitations are induced by external radiation fields and energy is dissipated through radiative decay, are examples of driven open quantum systems.
We explain the use of commutator-free exponential time-propagators for the 
numerical solution of the associated Schr\"odinger or master equations with a time-dependent Hamilton operator.
These time-propagators are based on the Magnus series
but avoid the computation of commutators, which makes them suitable  
for the efficient propagation of systems with a large number of degrees of freedom.
We present an optimized fourth order propagator and demonstrate its efficiency in comparison to the direct Runge-Kutta computation.
As an illustrative example we consider the parametrically driven dissipative Dicke model, 
for which we calculate the periodic steady state and the optical emission spectrum.
\end{abstract}

\maketitle

\section{Introduction}

The outcome of many experimental measurements is well described by linear response theory
for situations close to thermal equilibrium.
Other experiments, predominantly those dealing with small quantum systems in strong external fields, require a full non-equilibrium description.
One example is cavity quantum electrodynamics~\cite{WVEB06}, and generally finite quantum systems in radiation fields.
While the interaction of a single atom or an atomic ensemble with the quantized cavity field is weak,
transitions between atomic levels can be induced with strong, classical laser fields.
Through cavity losses and spontaneous emission the energy input from the external pumping
dissipates. Atoms in a cavity are open quantum systems far from equilibrium.

Such situations are described either by the Schr\"odinger equation
\begin{equation}\label{Schroeder}
\ii \partial_t |\psi(t)\rangle = H(t) |\psi(t)\rangle
\end{equation}
with a time-dependent Hamilton operator $H(t)$ if we neglect dissipation,
or more generally by a master equation
\begin{equation}\label{Master}
\partial_t \rho(t) = \mathcal{L}(t) \rho(t)
\end{equation}
with a time-dependent Liouville operator $\mathcal{L}(t)$,
e.g. one of Lindblad-type which includes dissipation in the Markovian approximation~\cite{BP02}.
In addition to single-time expectation values, which provide the basic information
from time propagation of the wave function $|\psi(t)\rangle$ or density matrix $\rho(t)$, one is interested in many-time correlations functions that yield optical spectra or information about the coherence or statistical properties of the emitted light~\cite{Carm99,VW06}.

Since explicit solutions of linear differential equations with 
variable coefficients do not exist apart from simple situations,
the above equations fall into the domain of numerical time-propagation.
The topic of the present paper is the application of commutator-free propagators based on the Magnus series~\cite{Ma54}. 
The Magnus series arises in the context of differential equations on Lie groups,
where it allows, among many other things, for the systematic construction of high-order approximations to the propagator~\cite{BCOR09}.
Commutator-free exponential time-propagators (CFETs)
avoid the use of commutators that appear in the Magnus series.
They provide an efficient and accurate algorithm for numerical time propagation~\cite{BM06},  
which we discussed for the Schr\"odinger equation in reference~\cite{AF11}.
Here, we concentrate instead on master equations for open quantum systems.
Although the present application lies outside of the principal Lie group setting,
we feel that the numerical results presented here are promising enough to warrant closer inspection.
The application to the parametrically driven dissipative Dicke model in section~\ref{sec:Dicke} gives an indication of the potential of this approach in non-trivial situations.

The paper is organized as follows.
In sections~\ref{sec:NumProb} and~\ref{sec:Magnus} we discuss the basic numerical problem and its principal solution through the Magnus series.
The commutator-free exponential time-propagators are introduced as a more practical solution in section~\ref{sec:CFET}, and an optimized 4th-order propagator is given in~\ref{sec:OptCFET}.
After a demonstration of their usage with the example of a spin in a magnetic field in section~\ref{sec:App} we turn to a discussion of the parametrically driven dissipative Dicke model in section~\ref{sec:Dicke}, before we summarize in section~\ref{sec:Summary}.

\section{The numerical problem}\label{sec:NumProb}

To describe the basic numerical problem we
consider the Schr\"odinger equation~\eqref{Schroeder}.
The standard approach obtains the wave function $|\psi(t)\rangle$ through time stepping.
The middle-point approximation
\begin{equation}\label{Middle}
  |\psi(t+\dt)\rangle \approx \exp \Big[-\ii \, \dt \, H(t+\dt/2) \Big] |\psi(t)\rangle
\end{equation}
allows for propagation of $|\psi(t)\rangle$ over a short time interval $[t,t+\dt]$.
Repeated application of equation~\eqref{Middle} gives $|\psi(t+T)\rangle$ starting from $|\psi(t)\rangle$ with $N=T/\dt$ time steps.
As detailed later, straightforward expansion of the exponential shows that the error of one time step with equation~\eqref{Middle} is $\propto (\dt)^3$,
such that the total error $\propto N (\dt)^3 = T (\dt)^2$ for propagation time $T$ scales as $(\dt)^2$.
Conversely, the achieved accuracy scales as $(\dt)^{-2}$,
which we can write symbolically as $\mathsf{error} \propto \mathsf{effort}^{-2}$.

As a second-order method the middle-point approximation is not efficient and requires small $\dt$ even for low accuracy demands.
If we ask for a better scheme we should note that the approximation~\eqref{Middle} has
two independent sources of error.
The genuine error in the situation of a time-dependent Hamilton operator
arises from the replacement of $H(t)$ by the constant $H(t+\dt/2)$,
and depends mainly on the rate of change of $H(t)$.
In addition, the numerical computation of an operator or matrix exponential $\exp[ A ]$ involves an error determined by the spread of eigenvalues of $A$.
In equation~\eqref{Middle} it is roughly proportionally to $\dt$ and the norm $\| H(t+\dt/2) \|$. 

Often the rate of change of $H(t)$, e.g. set by an external field frequency, is smaller than the largest eigenvalues of the Hamilton operator corresponding to highly excited states.
Then the total error is dominated by the computation of the exponential in equation~\eqref{Middle}.
Very small time steps $\dt$ and correspondingly large effort are required even if $H(t)$ changes slowly.

This observation explains why the use of general algorithms for the solution of ordinary differential equations (ODEs),
e.g. the standard 4th-order Runge-Kutta (RK4) procedure~\cite{PFTV86},
cannot be recommended unreservedly for the Schr\"odinger equation.
The problem of such (explicit) ODE solvers is that they provide only a poor approximation of the exponential in equation~(\ref{Middle}) and are inefficient already for constant $H$.

\begin{figure}
\includegraphics[width=0.48\textwidth]{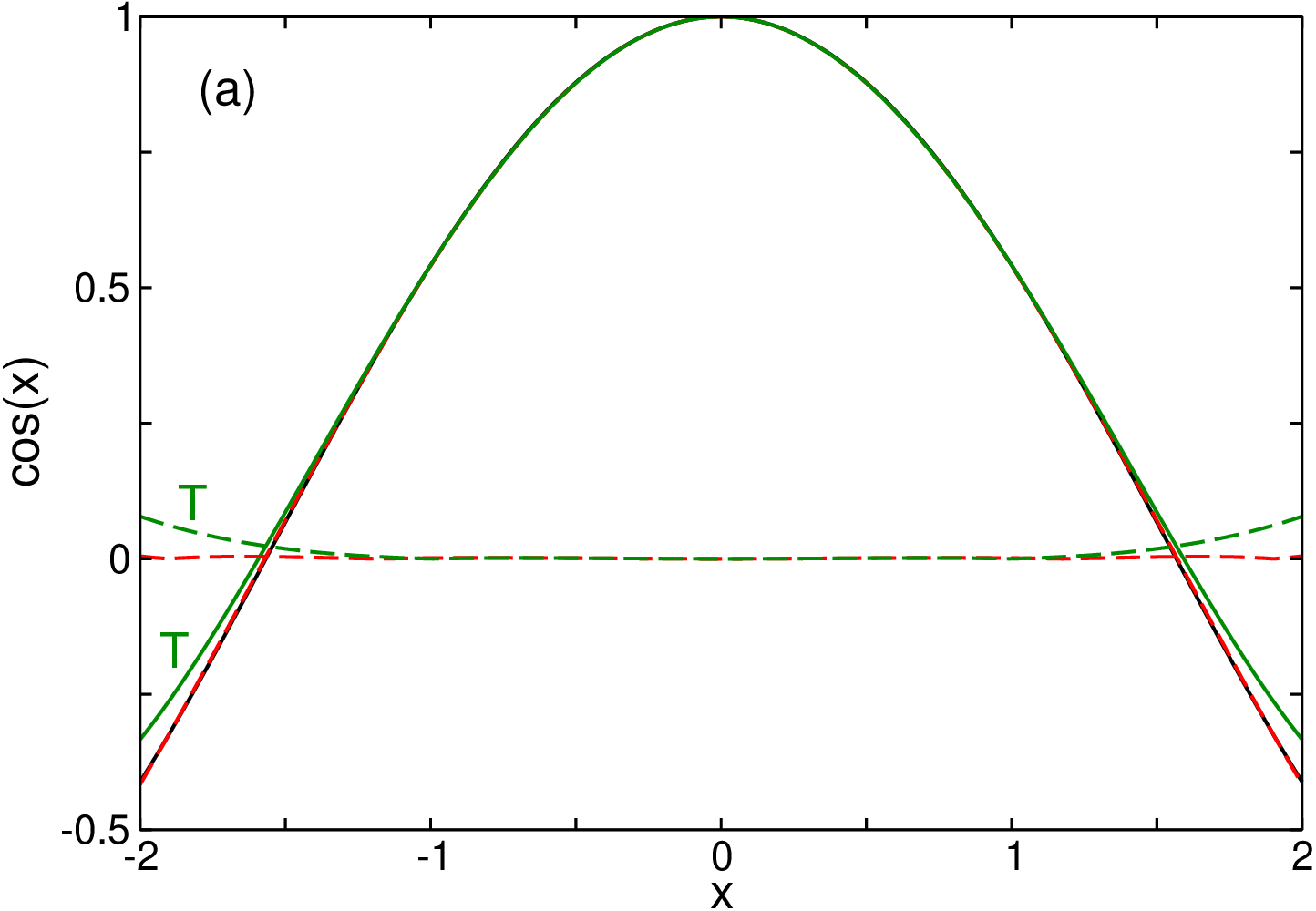}
\hfill
\includegraphics[width=0.48\textwidth]{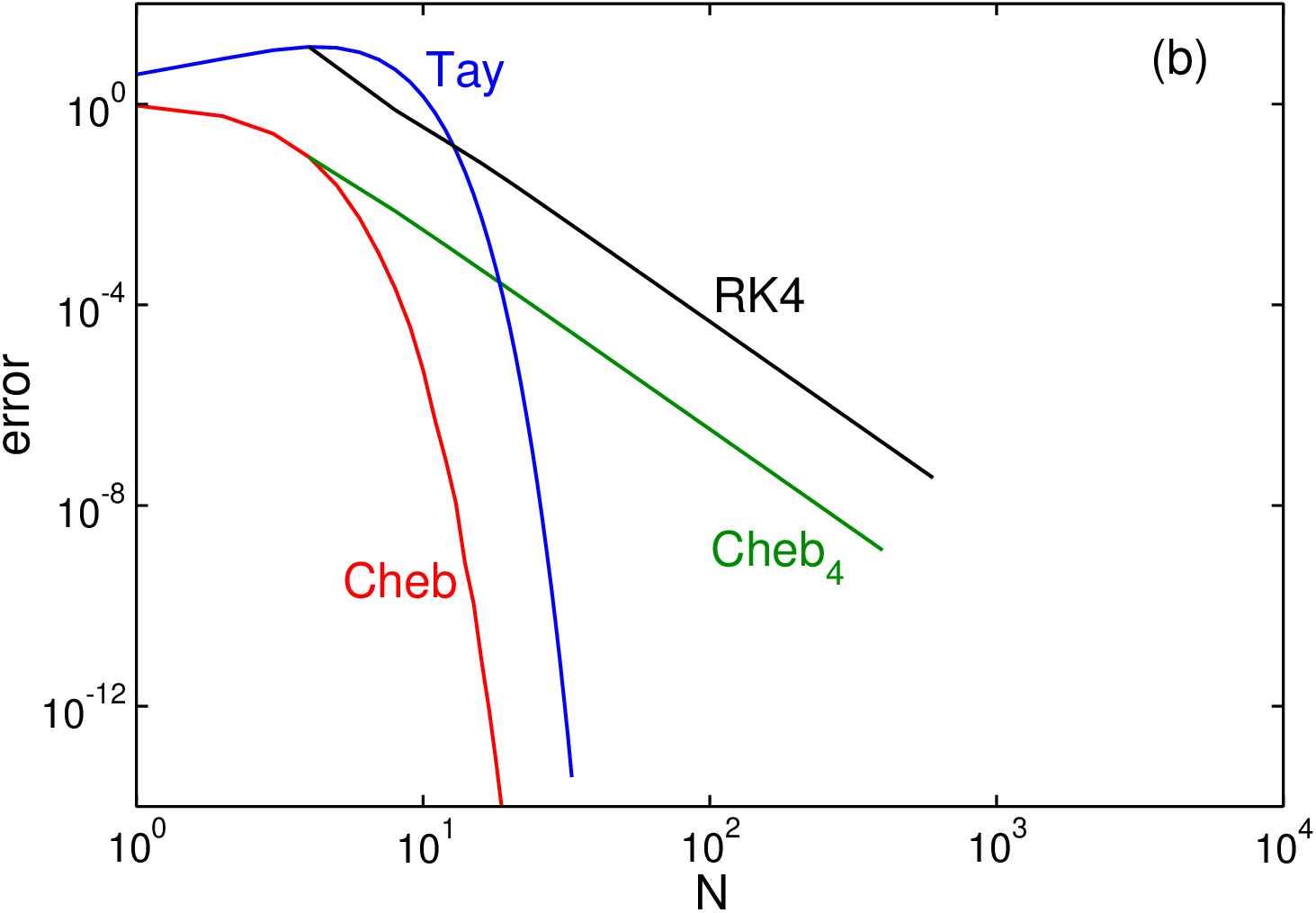}
\caption{Left panel (a): 4th-order Taylor (green curve) and Chebyshev approximation (red curve) of $\Re \exp(t) = \cos(t)$ on the interval $[-2,2]$. The dashed lines gives the error of both approximations.
The maximal error is $7.8 \times 10^{-2}$ for the Taylor approximation, which loses accuracy at the boundaries of the interval, versus $4.7 \times 10^{-3}$  for the Chebyshev approximation.
Right panel (b): Error $\varepsilon$ (see equation~\eqref{Error}) for the calculation of $\exp[-\ii H t]$ with the diagonal $10\times10$ matrix with eigenvalues $H_{nn}=n$ and $t=\pi/5$.
We compare the 4th-order Runge-Kutta procedure (RK4) with the use of the 4th-order Chebyshev approximation in a time-stepping scheme (Cheb$_4$),
and with a single propagation step using $N$ terms of the Taylor series (Tay) or of the Chebyshev approximation (Cheb).
In time-stepping the error decays as a power (here $\propto N^{-1/4}$) of the effort, while full computation of the exponential in a single step achieves much quicker error reduction.}
\label{fig:Exp}
\end{figure}

For example, the RK4 procedure approximates the exponential by the 5 terms 
$\exp[A] \approx 1 + A + A^2/2 + A^3/6 + A^4/24$ 
of the Taylor series of $e^x$.
The problem is that the Taylor series is not a good approximation unless $|x|$ is very small.
This effect is shown in panel (a) in figure~\ref{fig:Exp}, where it is compared to an approximation using Chebyshev polynomials (of the first kind)~\cite{Bo01}.
In this example, the five term Chebyshev approximation is $16$ times more accurate than the five term Taylor series.
For the 4th order approximation $\mathsf{error} \propto \mathsf{effort}^{-4}$, this implies that the efficiency is increased by a factor $16^{1/4}=2$.

In panel (b) in figure~\ref{fig:Exp} we compare the error-effort relation of the RK4 procedure to
that of the Chebyshev approximation for the calculation of $\exp[-\ii \dt H]$,
where $H$ is a diagonal matrix with entries $H_{nn}=n$ as for a harmonic oscillator. 
Here, and also in later examples (figures~\ref{fig:Spin1} and~\ref{fig:Spin2}),
we give the error between an exact and numerical matrix $A^{e/n}$ as the maximal difference of matrix elements
\begin{equation}\label{Error}
 \varepsilon = \max_{ij} \, |A^e_{ij}-A^n_{ij}| \; .
\end{equation} 
The Chebyshev approximation is clearly superior already for a small $\dt$.
It becomes even better with increasing eigenvalue spread or time-step $|\dt|$.

This reasoning motivates the replacement of ODE solvers by techniques which take advantage of the linearity of the Schr\"odinger or master equations.
The calculation of a matrix exponential is better accomplished with specialized algorithms,
such as split-operator methods~\cite{MQ02},
or the Krylov~\cite{PL86,HL97} or Chebyshev technique~\cite{TK84} in the case of large sparse matrices.
They allow for efficient propagation with time-independent Hamilton operators.
Equipped with such algorithms it remains to improve on the genuine error $\propto (\dt)^2$ involved in equation~\eqref{Middle} when turning to time-dependent Hamilton operators. 

\section{Magnus propagators}\label{sec:Magnus}

The importance of accurate evaluation of exponentials for the Schr\"odinger equation is related to the fact that the exponential maps the hermitian Hamilton operator $H$ onto the unitary propagator $\exp[-\ii t H ]$. Many differential equations involve a Lie algebra (here: of hermitian Hamiltonians) and a Lie group (here: of unitary propagators) in this way.
The idea behind ``geometric numerical integration'' of ODEs~\cite{HLW06}
is that also an approximate propagator should stay in the respective Lie group.

Let us consider  general linear differential equations
\begin{equation}\label{DGL}
 \dot{x}(t) = A(t) x(t) \;,
\end{equation}
where $x(t)$ is a vector and the coefficient matrix $A(t)$ is time-dependent.
The formal solution of equation~\eqref{DGL} is provided by the propagator $U(t)$
that gives
\begin{equation}
    x(t) = U(t) x(0) 
\end{equation}
for all $x(0)$.

For a scalar equation $\dot{x} = a(t) x$, the propagator is obtained through integration $U(t) = \exp[\int_0^t a(\tau) d\tau]$.
For operators or matrices $[A(t_1),A(t_2)] \ne 0$ is possible, such that this expression does not generalize.
We can still read this expression as an approximation
\begin{equation}\label{BeforeMagnus}
 U(t) \approx \exp \Big[  \int_0^t A(\tau) d\tau \Big] \;.
\end{equation}
For small $t=\dt$, this is nothing else than the approximation~(\ref{Middle}),
if the integral over $\tau$ is approximated (also with error $(\dt)^3$) using the middle-point value $A(\dt/2)$.
Although equation~\eqref{BeforeMagnus} involves a finite, maybe large, error
its exponential form guarantees that the approximate propagator lies in the Lie group.
The question  is whether we can improve on the $(\dt)^3$ scaling of the error and preserve the exponential form.

An affirmative answer is given by the Magnus series~\cite{Ma54,BCOR09}, which gives
\begin{equation}\label{Magnus}
 U(t) = \exp \Big[  \int_0^t A(\tau) d\tau  + \Omega_2(t) + \Omega_3 (t) + \dots  \Big] 
\end{equation}
as an exponential of Lie algebra elements $\Omega_n(t)$ and provides a systematic scheme for their construction.

The first term in~\eqref{Magnus} is the term known from the scalar case.
The non-commutativity of $A(t)$ is accounted for by 
correction terms $\Omega_n(t)$, for $n \ge 2$.
The term $\Omega_n(t)$ is given by a time-ordered integral of $n$-fold nested commutators of $A(\tau_i)$ and can be obtained through a recursive calculation.
The first two terms are
\begin{equation}
 \Omega_2(t) = \frac{1}{2} \int_0^t d\tau_1 \int_0^{\tau_1} d\tau_2 \, [A(\tau_1),A(\tau_2)] 
 \end{equation}
 and 
 \begin{equation}
 \fl
 \qquad  \Omega_3(t) = \frac{1}{6} \int_0^t \! d\tau_1 \int_0^{\tau_1} \! d\tau_2 \int_0^{\tau_2} \! d\tau_3 \, [A(\tau_1),[A(\tau_2),A(\tau_3)]] + [[A(\tau_1),A(\tau_2)],A(\tau_3)]  \;.
 \end{equation}
Explicit expressions for higher-order terms become quickly unwieldy.
Importantly, by building terms from commutators of Lie algebra elements $A(\tau_i)$, every $\Omega_n(t)$ stays in the Lie algebra.

The Magnus series solves two problems.
On the one hand, it preserves the Lie group structure of a differential equation.
 For the Schr\"odinger equation, where $A(t)= -\ii H(t)$,
 the propagator $U(t)$ is unitary as the exponential of an anti-hermitian matrix. Furthermore, unitarity of $U(t)$ is preserved for any truncation of the Magnus series.
On the other hand, since the term $\Omega_n(t)$ involves an $n$-fold integration over time,
its size scales as $(\dt)^{n}$.
Working with a truncated series including terms $\Omega_n$ for $n \le N$ only,
the error of the obtained approximate propagator itself scales as $(\dt)^{N+1}$.
We can thus improve systematically on the middle-point approximation~\eqref{Middle} by including more terms from the Magnus series.

Unfortunately, the Magnus series does not solve the practical problem of finding an efficient numerical time-propagation algorithm.
Computation of the nested commutators and multiple integrals is difficult
to implement and consumes computational resources. 
Fortunately, there is a simpler and more convenient way.

\section{Commutator-free exponential time-propagators}\label{sec:CFET}

The use of commutator-free exponential time-propagators (CFETs) has been discussed in references~\cite{BM06,Thal06,AF11}. They are, basically, a reformulation of the Magnus series that avoids integrals and commutators and gives the propagator as a product of exponentials of simple linear combinations of $A(t)$.

The simplest CFET is the middle-point approximation~\eqref{Middle} itself.
A 4th-order CFET, where the error scales as $(\dt)^4$, was introduced in~\cite{BM06,Thal06}.
It gives the approximate propagator as the product of two exponentials
\begin{equation}\label{SimpleCFET}
   U_\mathrm{CFET}(\dt) = \exp \Big[ \dt \left( g_1 A^{(1)} + g_2  A^{(2)} \right) \Big] \,
                                    \exp \Big[ \dt \left( g_2 A^{(1)} + g_1  A^{(2)} \right) \Big] \;, 
\end{equation}
which involve a linear combination of $A(t)$
specified by the coefficients
\begin{equation}\label{GSimple}
 g_1 = \frac{3-2\sqrt{3}}{12} \;, \quad
 g_2 = \frac{3+2\sqrt{3}}{12} \;,
 \end{equation}
and uses only the values 
\begin{equation}
 A^{(1)} = A\big[  x_1 \dt \big] \;, \quad
 A^{(2)} = A \big[ x_2 \dt \big] 
\end{equation}
of $A(t)$ evaluated at two points in $[0,\dt]$ given by
\begin{equation}\label{XSimple}
 x_1 = \frac{1}{2}-\frac{\sqrt{3}}{6} \;, \quad
 x_2 = \frac{1}{2}+\frac{\sqrt{3}}{6} \;.
\end{equation}
 
This expression is the simplest non-trivial CFET.
A better, optimized, 4th-order CFET is presented below in equation~\eqref{CFET}.
Higher-order CFETs can be constructed, and the freedom in the choice of coefficients can be exploited for their optimization, i.e. the minimization of the error.
The construction of CFETs is rooted in the theory of abstract free Lie algebras that underlies the Magnus series.
Its description is beyond the scope of this paper,
and we refer the reader to reference~\cite{BM06} and our reference~\cite{AF11} for details.
Here, we proceed in the opposite way and give a direct check of the validity of equation~\eqref{SimpleCFET} that avoids most of the language of free Lie algebras.

\subsection{Direct validation of the 4th-order CFET}

The principle idea is to combine the two exponentials in equation~\eqref{SimpleCFET} with the Baker-Campbell-Hausdorff (BCH) formula $\exp [X] \exp[Y] = \exp[X+Y + [X,Y]/2 + \dots ]$ and compare the resulting expression with the original Magnus series.
Let us begin with the Taylor series
\begin{equation}
  A(t) = A_1 + A_2 t + A_3 t^2 + A_4 t^3 + O (t^4)
\end{equation}
of $A(t)$, in the vicinity of $t=0$.
For the 4th-order CFET, only terms $A_1, \dots, A_4$ have to be considered.

We insert the Taylor series in the Magnus series~\eqref{Magnus}, and keep the first three terms to $\Omega_4(t)$.
The terms $\Omega_n(t)$ for $n \ge 5$ give contributions of order $(\dt)^5$ and higher.
A simple counting of indices shows that only the seven terms
$A_1, A_2, A_3, A_4, [A_1, A_2], [A_1,A_3], [A_1,[A_1,A_2]]$
can contribute in fourth order.
The Magnus series thus gives the propagator
\begin{eqnarray}\label{MagnusExplicit}
\fl
\quad U(\dt) = \exp \bigg[ & \dt A_1 + \frac{(\dt)^2}{2} A_2    + (\dt)^3 \Big( \frac{1}{3} A_3 - \frac{1}{12} [A_1,A_2] \Big)   \\
 & +  (\dt)^4 \Big( \frac{1}{4} A_4 - \frac{1}{12} [A_1,A_3]  \Big) + O\Big((\dt)^5\Big) \bigg] \;. \nonumber
\end{eqnarray}
Note that the commutator $[A_1,[A_1,A_2]]$ does not contribute.
This expression is the exact reference for comparison.

It is easy to see that the middle-point approximation~\eqref{Middle} is correct to second order $(\dt)^2$:
The terms $\dt A_1 + ((\dt)^2/2) A_2$ are reproduced exactly,
but the commutator $[A_1,A_2]$ in the third order term is missing.

For the 4th order CFET from equation~\eqref{SimpleCFET}, it is
\begin{equation}
 A^{(k)}  =  A_1 + \dt \, x_k \, A_2 + (\dt \, x_k)^2 \, A_3 + (\dt \, x_k)^3 \, A_4  + O((\dt)^4) 
\end{equation}
for $k=1,2$, which inserted gives
\begin{eqnarray}\fl
     U_\mathrm{CFET}(\dt)  = &  \quad \exp \Big[ \dt (g_1+g_2) A_1 + (\dt)^2 (g_1 x_1 + g_2 x_2) A_2 \\ 
     & \qquad\qquad  + (\dt)^3 (g_1 x_1^2 + g_2 x_2^2) A_3 + (\dt)^4 (g_1 x_1^3 + g_2 x_2^3) A_4 + O((\dt)^5) \Big] \nonumber \\[1ex]  
     & \times  \exp \Big[ \dt (g_2+g_1) A_1 + (\dt)^2 (g_2 x_1 + g_1 x_2) A_2 \nonumber\\
     & \qquad\qquad + (\dt)^3 (g_2 x_1^2 + g_1 x_2^2) A_3 + (\dt)^4 (g_2 x_1^3 + g_1 x_2^3) A_4 + O((\dt)^5) \nonumber \Big] \;.
\end{eqnarray}

We now use the BCH formula
\begin{equation}\label{BCH}
\fl \quad
e^X e^Y = \exp \Big[ 
X  + Y +\frac{1}{2} [X,Y] +\frac{1}{12} [X,[X,Y]] -\frac{1}{12} [Y,[X,Y]]
+ \dots \Big] 
\end{equation}
to combine the two exponentials.
Only the three commutators shown in equation~\eqref{BCH} need to be evaluated, the following commutators in the BCH formula contribute terms of order $(\dt)^5$ or higher.
We then obtain the expression
\begin{eqnarray}\label{CFETExplicit}
\fl
 U_\mathrm{CFET}(\dt) = \exp \Big[ & \dt  \xi_1 A_1 + (\dt)^2 \xi_2 A_2    + (\dt)^3 \Big( \xi_3 A_3 + \chi_1 [A_1,A_2] \Big)   \nonumber \\
 & +  (\dt)^4 \Big( \xi_4 A_4 + \chi_2 [A_1,A_3] + \chi_3 [A_1,[A_1,A_2]]   \Big) + O((\dt)^5) \Big] \;,
\end{eqnarray}
which allows for direct comparison with the Magnus series in equation~\eqref{MagnusExplicit}.
We immediately recognize the seven terms and commutators from there. 

Comparison of the coefficients $\xi_i$, which we evaluated with the BCH formula, to the prefactors in~\eqref{MagnusExplicit} gives the conditions
\numparts
\begin{equation}\label{CondA}
  \xi_1 = 2 g_1 + 2 g_2 = 1 \;,
\end{equation}
\begin{equation}\label{CondB}
  \xi_2  = (g_1+g_2) (x_1 + x_2) = \frac{1}{2} \;,
\end{equation}
\begin{equation}\label{CondC}
  \xi_3  = (g_1 + g_2) (x_1^2+x_2^2) = \frac{1}{3} \;,
\end{equation}
\begin{equation}\label{CondD}
  \xi_4  = (g_1 + g_2) (x_1^3+x_2^3) = \frac{1}{4} \;,
\end{equation}
\begin{equation}\label{CondE}
 \chi_1 = \frac{1}{2} (g_1+g_2) \Big( (g_2 x_1 + g_1 x_2) - (g_1 x_1 + g_2 x_2)  \Big) 
  =  - \frac{1}{12} \;,
 \end{equation}
\begin{equation}\label{CondF}
 \chi_2 = \frac{1}{2} (g_1+g_2) \Big( (g_2 x_1^2 + g_1 x_2^2) - (g_1 x_1^2 + g_2 x_2^2) \Big) =  - \frac{1}{12} \;,
 \end{equation}
\begin{equation}\label{CondG}
 \chi_3 = \frac{1}{12} (g_1+g_2)^2 \Big( (g_2 x_1 + g_1 x_2) - (g_2 x_1 + g_1 x_2)  \Big)   =  0 \;.
 \end{equation}
 \endnumparts
 To complete our check we insert $x_1, x_2$ from equation~\eqref{XSimple} and
 $g_1, g_2$ from~\eqref{GSimple} and find that all the seven conditions are satisfied. 
 Note that condition~\eref{CondG}, and also~\eref{CondD} and~\eref{CondF}, are redundant.

For the construction of a CFET, this process has to be reversed.
We start from an ansatz with a product of exponentials, derive the relevant conditions for a CFET of required order, and solve the resulting polynomials equations for possible coefficient values.
Several adjustments of the direct calculation done here simplify bookkeeping,
and reveal underlying structures which reduce the number of conditions.
Still, the construction of higher-order CFETs is involved and not entirely free of brute force computations.
We refer the reader to reference~\cite{AF11} to get an impression,
where CFETs up to order $8$ are presented.
The usage of a given CFET, however, is plain and simple.

\subsection{Optimized 4th order CFET}\label{sec:OptCFET}

For the application to dissipative systems we recommend here the use of an optimized 4th-order CFET, which is equation (43) in our reference~\cite{AF11}.
Extending the simpler expression~\eqref{SimpleCFET} it gives the approximate propagator as the product of three exponentials
\begin{eqnarray}\label{CFET}
 U^\mathrm{opt}_\mathrm{CFET}(\delta t) = & \quad \exp \Big[ \dt \left( g_1 A^{(1)} + g_2 A^{(2)} + g_3 A^{(3)} \right)  \Big]  \, \nonumber \\[0.5ex]
         & \times \exp\Big[  \dt \left( g_4 A^{(1)} + g_5 A^{(2)} + g_4 A^{(3)} \right)  \Big] \, \\[0.5ex]
         & \times \exp\Big[ \dt \left( g_3 A^{(1)} + g_2 A^{(2)} + g_1 A^{(3)} \right)  \Big]  \nonumber\;, 
\end{eqnarray}
where 
\begin{equation}
 A^{(1)} = A\big[  x_1 \dt \big] \;, \quad
 A^{(2)} = A \big[ x_2 \dt \big] \;,  \quad
 A^{(3)} = A\big[  x_3 \dt \big]
\end{equation}
with
\begin{equation}
 x_1 =  \frac{1}{2} - \sqrt{\frac{3}{20}}  \;, \quad
 x_2 = \frac{1}{2} \;, \quad
 x_3 = \frac{1}{2} + \sqrt{\frac{3}{20}}  \;,
 \end{equation}
and
\begin{eqnarray}
 g_1 = \frac{37}{240} - \frac{10}{87} \sqrt{\frac{5}{3}} \;, \quad
 g_2 = - \frac{1}{30} \;, \quad 
 g_3 = \frac{37}{240} + \frac{10}{87} \sqrt{\frac{5}{3}}  \;, \quad \nonumber \\[2ex]
 g_4 = -\frac{11}{360} \;, \quad
 g_5 = \frac{23}{45} \;.
\end{eqnarray}

The error of this CFET scales again as $(\dt)^5$, but the prefactor in front of the error term is considerably smaller than for the CFET~\eqref{SimpleCFET}. 
The reduction outweighs the increase of effort using three instead of two exponentials.

In contrast to the original Magnus series usage of the CFET~\eqref{CFET} is compellingly easy.
Only linear combinations of $A(t)$ evaluated at three different points in the interval $[0,\dt]$ need to be formed. All commutators and integrations have been removed from the expression.
Of course, we assume the existence of an algorithm for the computation of matrix exponentials.

Let us stress the advantage of easy usage with the even simpler formulation that is obtained for an $A(t) = B + f(t) C$ (it is easily generalized to include more terms).
In this case equation~\eqref{CFET} can be written as
\begin{equation}\label{CFET2}
\fl
 U_\mathrm{CFET}(\delta t) =  \exp \Big[ \dt_1 \left( B + f_1 C \right)  \Big]  \, \exp\Big[  \dt_2 \left( B + f_2 C \right)  \Big] \, \exp\Big[ \dt_1 \left( B + f_3 C \right)  \Big] \;, 
\end{equation}
with the time steps
\begin{equation}
 \dt_1 = \frac{11}{40} \dt \;, \quad  \dt_2 = \frac{9}{20} \dt \;,
\end{equation}
and the coefficients $f_1, f_2, f_3$ from
\begin{equation}\label{FCoeff}
\left(\begin{array}{c} f_1 \\[0.5ex] f_2 \\[0.5ex] f_3 \end{array}\right) =
\left( \begin{array}{ccc}
  h_1 & h_2 & h_3 \\[0.5ex]
  h_4 & h_5 & h_4 \\[0.5ex]
  h_3 & h_2 & h_1 
 \end{array}
 \right)
\left(\begin{array}{c} f(x_1 \dt) \\[0.5ex] f(x_2 \dt)  \\[0.5ex] f(x_3 \dt) \end{array}\right) \;,
\end{equation}
 using 
\begin{eqnarray}
 h_1 = \frac{37}{66}- \frac{400}{957} \sqrt{\frac{5}{3}} \;,
 \quad
 h_2 =  - \frac{4}{33} \;,
 \quad
 h_3 = \frac{37}{66} + \frac{400}{957} \sqrt{\frac{5}{3}} \;, \nonumber\\[0.2ex]
 h_4 = - \frac{11}{162} \;, 
 \quad
 h_5 =  \frac{92}{81} \;.
 \end{eqnarray}

Through the CFET the full propagation with a time-dependent term $f(t) C$ is replaced
(approximately) by piecewise propagation with a constant term $f_i C$.
The choice of the coefficients $f_1, f_2, f_3$ according to equation~\eqref{FCoeff} guarantees that the error of this approximation scales as $(\dt)^5$.
This is the signature of geometric integration~\cite{HLW06}: Figuratively speaking, instead of moving along a curve in the Lie group we move repeatedly along short straight lines, the direction of which is given by the linear combinations in equation~\eqref{CFET} or~\eqref{CFET2}.

Note that the time steps $\dt_1$, $\dt_2$ are positive
such that propagation proceeds in the forward direction.
This is important for dissipative systems where a negative $\dt_i$ would push eigenvalues of $\mathcal{L}(t)$ into the right half complex plane, which leads to exponentially growing terms and corresponding numerical instabilities.
For this reason we restrict ourselves to 4th-order CFETs here.

\subsection{Exemplary application of the optimized 4th-order CFET}\label{sec:App}

Let us apply the 4th-order CFET~\eqref{CFET} to a simple example and compare with the RK4 procedure which, as we claimed, should be less efficient because it does not properly compute exponentials.
We have to stress that the efficiency of CFETs depends on 
a good algorithm for the computation of matrix exponentials. 
Otherwise, when the additional computational overhead involved exceeds the savings achieved with a large time-step $\dt$, the simple Runge-Kutta procedure is more efficient.

 Good algorithm for the symmetric case ($A^\dagger = \pm A$, e.g. for a hermitian Hamiltonian)
 are the Chebyshev, Krylov and split-operator techniques mentioned before.
 For the unsymmetric case ($[A,A^\dagger] \ne 0$) encountered for dissipative systems, all techniques meet problems which are only partially solved.
 After suitable modifications of the standard procedure the Chebyshev technique behaves most favourably.
 The exploration of this point has to be left for a future publication,
here we take the virtues of the Chebyshev technique for granted.

\subsubsection{First example: Spin in a magnetic field}

As an example for a non-dissipative system consider a spin (length $j$) in a rotating magnetic field,
with Hamilton operator
\begin{equation}
 H(t) = 2 \Delta J_z + 2 V \cos 2 \omega t J_x + 2 V \sin 2 \omega t J_y \;.
\end{equation}
The time-evolution of the wave function can be determined exactly 
after transformation with $\exp[\ii \omega t J_z]$ to the rotating frame (see below). 

In figure~\ref{fig:Spin1} we plot the error-effort relation for the optimized CFET~\eqref{CFET} and the RK4 procedure for a typical set of parameters (the behavior for other parameters is identical).
The error $\varepsilon$ is determined as in equation~\eqref{Error}.
For the effort $N_H$ we count the number of evaluations of matrix-vector multiplications with the Hamilton operator, which is generally the most time-consuming step.
The matrix exponentials needed for the CFET are calculated with the Chebyshev technique to machine precision.

We see that the RK4 procedure requires lesser effort for low accuracy only, 
but use of the CFET becomes quickly advantageous as the spin length or propagation time increases. For $j=10$ and $t=100$ in panel (b), 
the CFET is more efficient for error goals less than $1\%$, with an a efficiency gain of a factor $2 \dots 4$.

\begin{figure}
\includegraphics[width=0.48\textwidth]{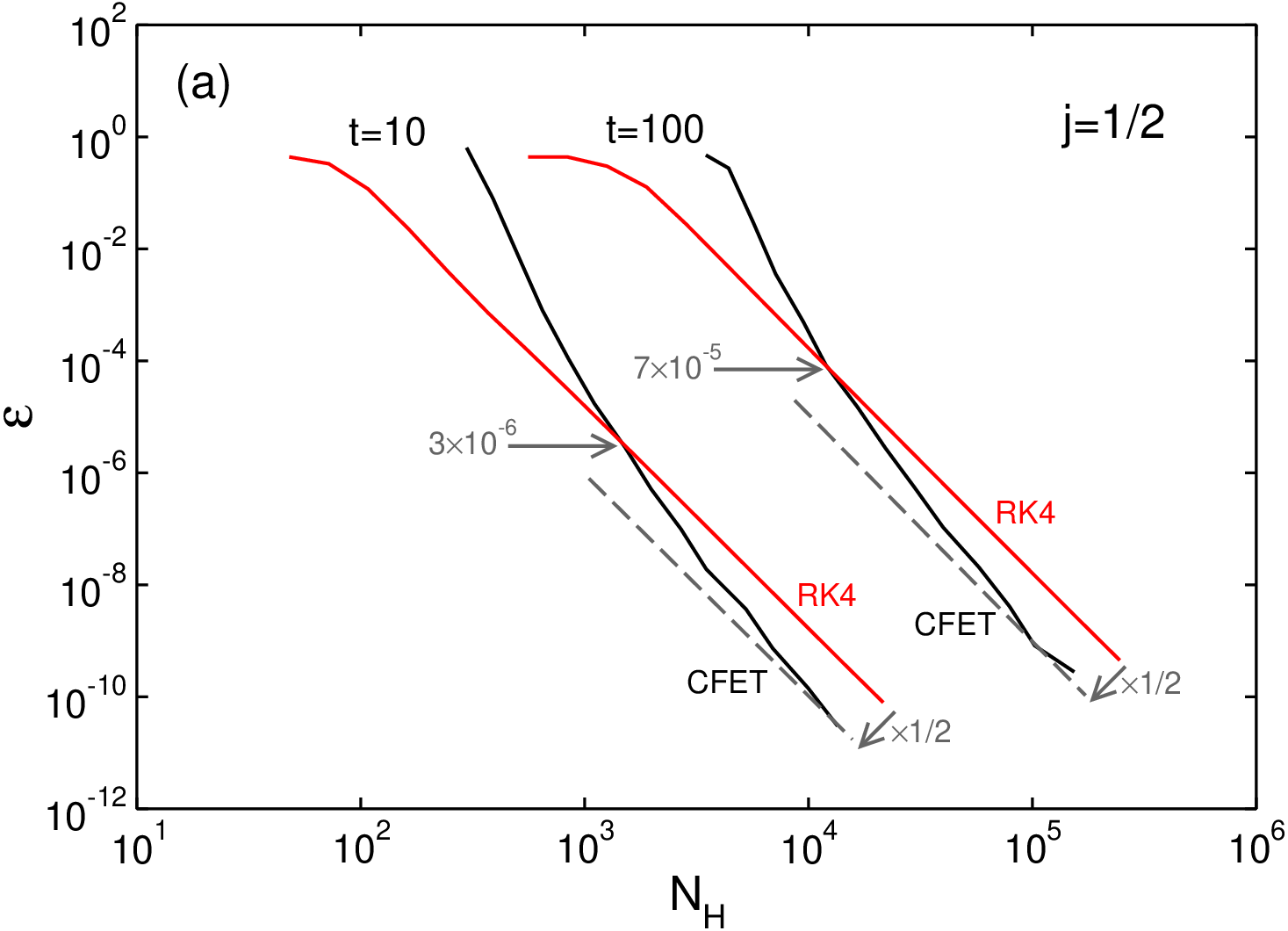} \hfill
\includegraphics[width=0.48\textwidth]{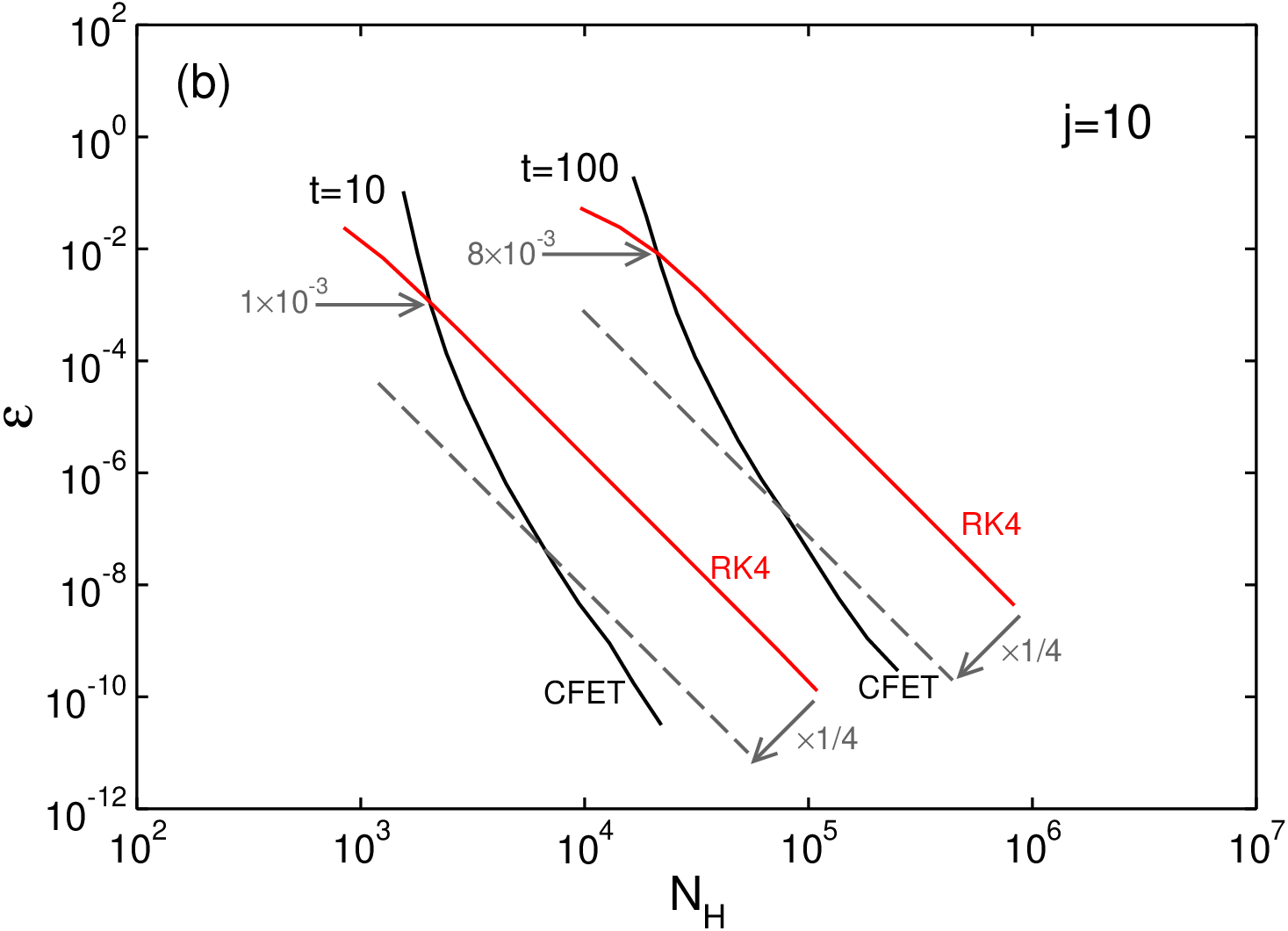} 
\caption{Error $\varepsilon$ between exact and numerical density matrices $\rho_{e/n}(t)$ (see equation~\eqref{Error}) versus the effort $N_H$ (number of matrix-vector multiplications with $H$) 
for $\Delta=V=\omega=1.0$
and $j=1/2$ (left panel (a)) or $j=10$ (right panel (b)).
The black curve gives the results using CFETs with a Chebyshev evaluation of the exponential,
the red curve using the RK4 procedure.
Curves are shown for propagation time $t=10$ and $t=100$.
The grey dashed lines indicate the reduction of the RK4 effort by a factor $1/2$ or $1/4$
achieved by the CFETs.
} 
\label{fig:Spin1}
\end{figure}

\subsubsection{Second example: Driven dissipative two-level system}

We keep the spin in the rotating magnetic field as an example and include dissipation.
With dissipation, its time evolution is described by a master equation~\eqref{Master} for the spin density matrix $\rho(t)$.

In the Lindblad formalism, the Liouville operator $\mathcal{L}(t) = \mathcal{L}_H(t) + \mathcal{L}_D$ is the sum of two or more terms with different meaning~\cite{BP02}.
The first term
\begin{equation}
 \mathcal{L}_H(t) \rho = - \ii [H(t),\rho]
\end{equation}
contains the Hamilton operator $H(t)$ and will be time-dependent.
The second and further terms have the form 
\begin{equation}\label{LindbladD}
 \mathcal{D}[A] \rho = 2 A \rho A^\dagger - A^\dagger A \rho -  \rho A^\dagger A \;.   
\end{equation}
They introduce eigenvalues of $\mathcal{L}(t)$ with finite (negative) real part
and thus describe dissipation.
Within the Lindblad formalism the form of $\mathcal{D}[A]$ guarantees that the structural properties of the density matrix -- hermiticity, normalization, positive semi-definiteness -- are strictly preserved.
Note that the numerical time propagation scheme does not depend on the precise form of $\mathcal{L}(t)$, as long as the master equation remains linear and local in time.

For the driven spin, dissipation is included through the Lindblad term
\begin{equation}
\mathcal{D}[J_-] \rho =  2  J_- \rho J_+   - J_+ J_- \rho - \rho J_+ J_- \;,
\end{equation}
and the full Liouville operator is
\begin{equation}\label{spinL}
 \mathcal{L} = \mathcal{L}[H] + \gamma \mathcal{D}[J_-] 
\end{equation}
with the dissipation rate $\gamma > 0$.

For $j=1/2$, the exact solution of this problem is possible with a transformation 
$\tilde{\rho}(t) = \exp[ \ii \omega t J_z  ] \rho(t) \exp[ -  \ii \omega t J_z   ]$
to the rotating frame, which gives a time-independent Hamilton operator 
$\tilde{H} = 2 (\Delta - \omega) J_z + 2 V J_x$.
Note that the transformation leaves $J_z$ invariant.

The stationary state in the rotating frame, corresponding to the eigenvalue zero of the transformed Liouville operator $\tilde{\mathcal{L}}$ for $\gamma >0$, is 
\begin{equation}
\fl
 \tilde{\rho}_\infty = \frac{1}{4 (\Delta-\omega)^2 + \gamma^2 + 2 V^2} \left(\begin{array}{cc}
V^2 & - (2 (\Delta-\omega) + \ii \gamma) V    \\[1ex]
- (2 (\Delta-\omega) - \ii \gamma) V
 &  4 (\Delta-\omega)^2 + \gamma^2 + V^2
   \end{array}\right) \;.
\end{equation}
In particular, $\langle J_z(t)\rangle$ converges  for $t \to \infty$, with constant value
\begin{equation}\label{JzSteady}
\langle J_z \rangle_\infty = \frac{V^2}{4 (\Delta-\omega)^2 + \gamma^2 + 2 V^2} - \frac{1}{2} \;.
\end{equation}

\begin{figure}
\includegraphics[width=0.48\textwidth]{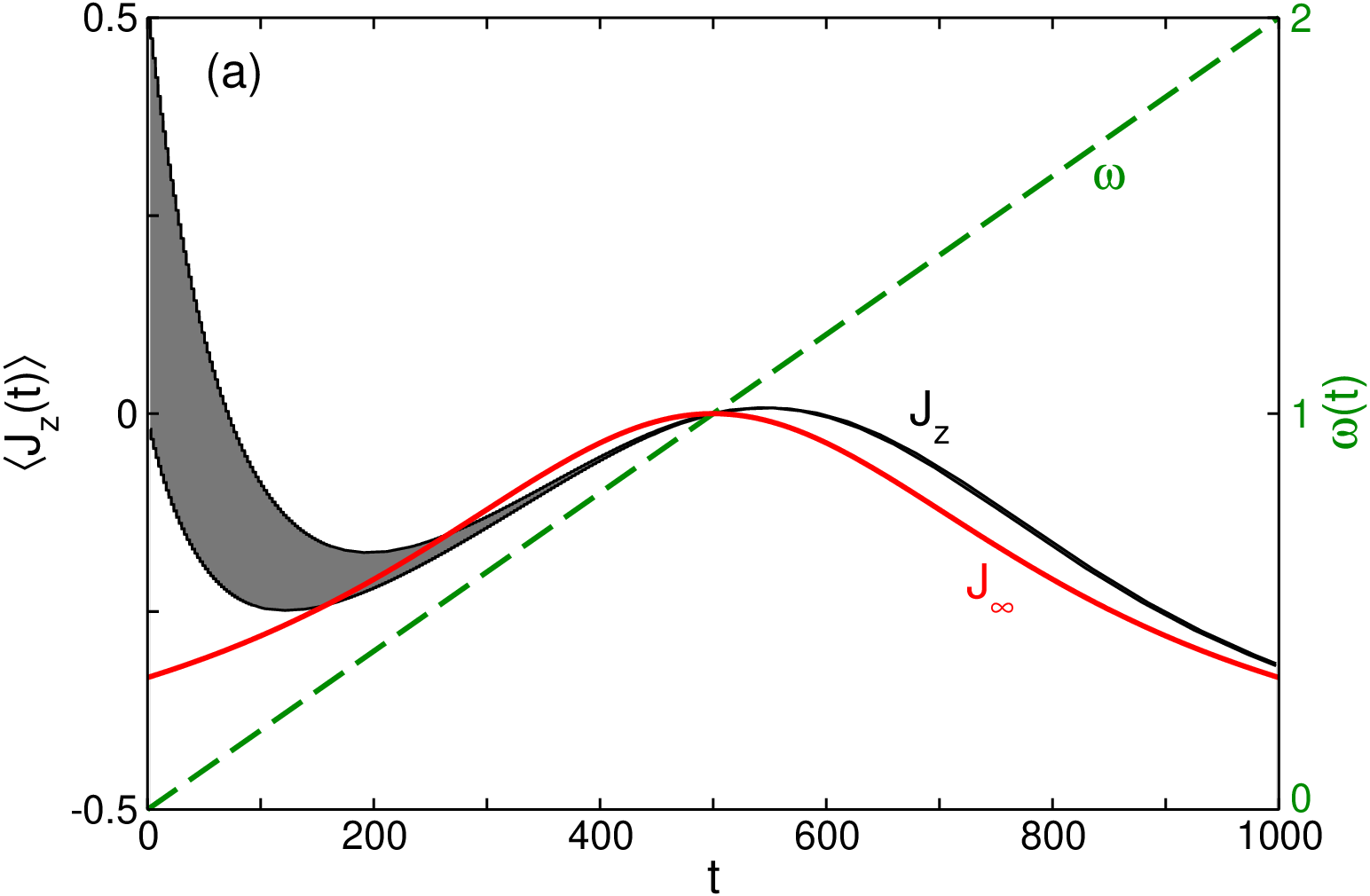}
\hfill
\includegraphics[width=0.48\textwidth]{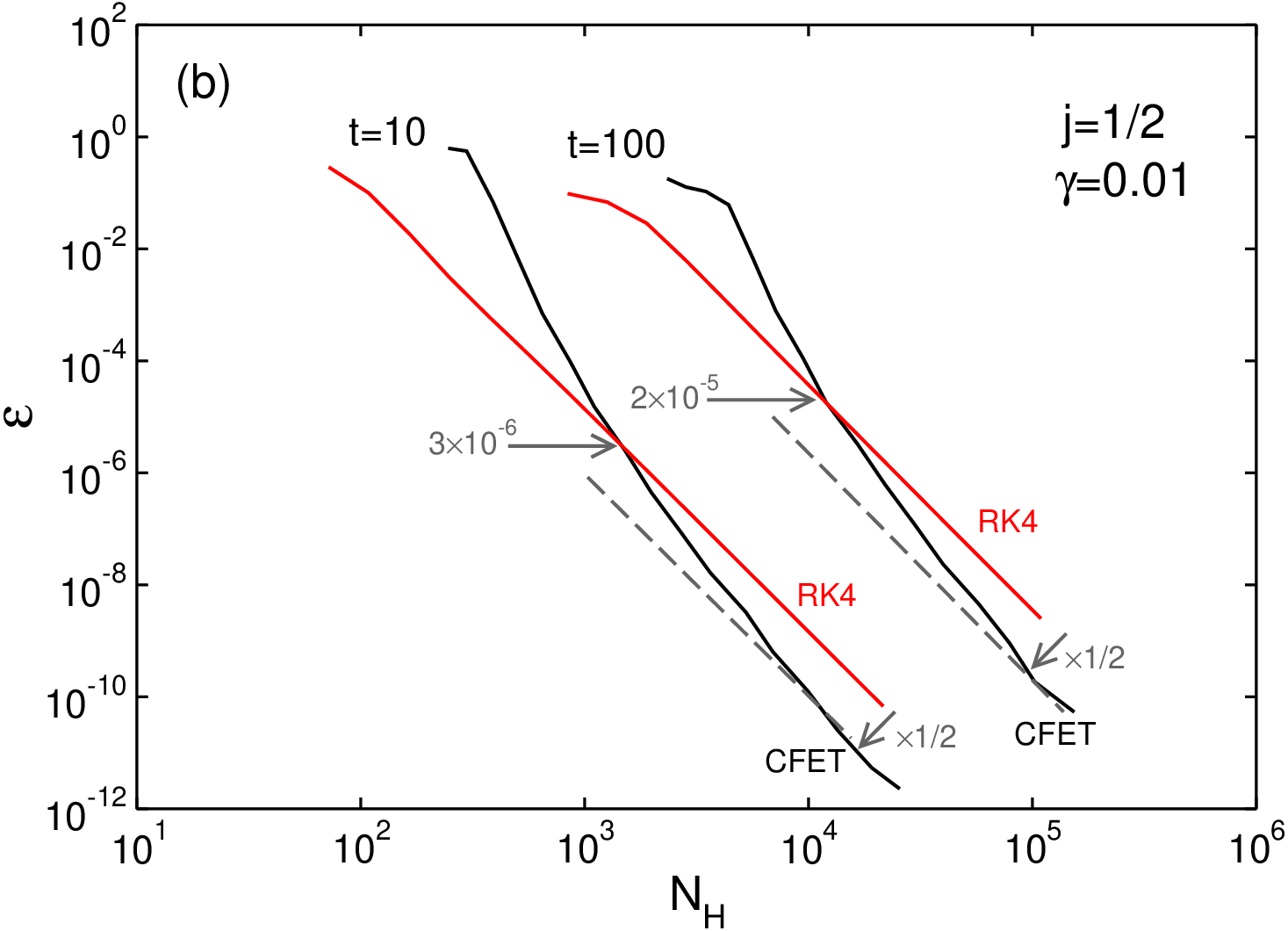}
\caption{Left panel (a): $\langle J_z(t) \rangle$ of the driven dissipative spin $j=1/2$ from equation~\eqref{spinL}, with $\Delta=V=\omega=1.0$ and $\gamma=0.01$.
Shown is the envelope function, suppressing the fast spin oscillations with frequency $\Delta$.
The magnetic field frequency $\omega(t)$ grows linearly from $0$ to $2$ during propagation (green dashed curve).
 The red curve gives the steady state result from equation~\eqref{JzSteady}
 to the respective frequency $\omega(t)$.
Right panel (b): Similar to figure~\ref{fig:Spin1}, the error $\varepsilon$ vs. effort $N_H$ for $\Delta=V=\omega=1.0$, $j=1/2$ and finite dissipation $\gamma=0.01$.
}
\label{fig:Spin2}
\end{figure}

In panel (a) in figure~\ref{fig:Spin2} we plot $\langle J_z(t) \rangle$ starting from the initial state with $\langle J_z(0) \rangle = +1/2$. Transient oscillations decay as a result of finite dissipation $\gamma >0$. The value of $\langle J_z(t)\rangle$ in the quasi-equilibrium state depends on the field frequency $\omega$, which we increase slowly during time-propagation.
$\langle J_z(t)\rangle$ follows closely the value $\langle J_z \rangle_\infty$ from equation~\eqref{JzSteady}, with a short delay,
and we identify the resonance at $\omega(t)= \Delta$.

At resonance $\omega = \Delta$, we get simple expressions for the remaining three eigenvectors of $\tilde{\mathcal{L}}$.
The non-zero eigenvalues are $\lambda_1 = - \gamma$, $\lambda_2 = - (3\gamma + \xi)/2$,
and $\lambda_3 = - (3 \gamma - \xi)/2$,
with corresponding eigenvectors
\begin{equation}
 \rho_1 = \left(\begin{array}{cc} 0 & 1 \\ 1 & 0  \end{array}\right) \;,
\end{equation}
\begin{equation}
 \rho_2 = \left(\begin{array}{cc} 1 & - \ii (\gamma-\xi)/(4V) \\ \ii (\gamma-\xi)/(4V) & -1  \end{array}\right) \;,
\end{equation}
\begin{equation}
 \rho_3 = \left(\begin{array}{cc} 1 & - \ii (\gamma+\xi)/(4V) \\ \ii (\gamma+\xi)/(4V) & -1  \end{array}\right) \;.
\end{equation}
We have introduced the abbreviation $\xi = \sqrt{\gamma^2 - 16 V^2}$.

Starting from the initial state with $\langle J_z(0) \rangle = \frac{1}{2}$,
we obtain the time evolution of $\tilde{\rho}(t)$ from the decomposition
\begin{equation}\fl
\rho(0)=  \left(\begin{array}{cc} 1 & 0 \\ 0 & 0  \end{array}\right)
= \rho_\infty + \frac{V^2+\gamma^2}{2(2V^2+\gamma^2)} (  \rho_2 + \rho_3 )
+ \frac{\gamma^3+ 5 \gamma V^2}{2 \xi (2 V^2 + \gamma^2)} (\rho_2 - \rho_3)
\end{equation}
of the density matrix.
In the underdamped case $\gamma<4 V$, we have
\begin{eqnarray}\label{JZSpin}
\fl\quad
 \langle J_z(t) \rangle =&  \frac{-\gamma^2}{2(\gamma^2+2 V^2)} \nonumber \\
 & + \frac{1}{2V^2+\gamma^2}  \left( (V^2+\gamma^2) \cos \tilde{\omega} t \;
 - (\gamma^3+ 5 \gamma V^2 ) \frac{\sin \tilde{\omega} t}{2 \tilde{\omega}} \right) e^{-(3/2)\gamma t} \;,
\end{eqnarray}
where we write $\tilde{\omega} = (1/2) \sqrt{16 V^2 - \gamma^2}$ for the frequency of the transient contribution.
These expressions allow for comparison with numerical results.

For the numerical solution of this problem, we do not transform the problem but keep the time-dependence of $H(t)$ explicitly.
In panel (b) in figure~\ref{fig:Spin2} we plot the error-effort relation as in figure~\ref{fig:Spin1}.
The relation between the CFET and the RK4 procedure is similar to the non-dissipative case, and we recognize the factor $1/2$ of error reduction for $j=1/2$.
The advantage of the CFET is however not quite as distinct as for the dissipation-free case in figure~\ref{fig:Spin1}, panel (a).

\section{The parametrically driven dissipative Dicke model}\label{sec:Dicke}

The previous examples serve as a benchmark for the CFET approach.
We can now demonstrate its usefulness for a less academic case,
the parametrically driven dissipative Dicke model.
Optical properties of the Rabi case $j=1/2$, corresponding to a single qubit, 
have been explored in reference~\cite{LGCC09},
and quantum phase transitions in the parametrically driven Dicke model without dissipation are studied in~\cite{BERB12}.

The Hamilton operator of the Dicke model~\cite{Di54},
\begin{equation}
 H(t) = \Delta J_z + \Omega a^\dagger a + \lambda(t) (a+a^\dagger) J_x  \;,	
\end{equation}
describes an ensemble of two-level atoms (transition energy $\Delta$) as a pseudo-spin of length $j$, which couples to the cavity field (frequency $\Omega$).
We assume a time-dependent interaction constant
\begin{equation}
 \lambda(t) = \lambda_0 + \delta \lambda \cos \omega_p t \;,
 \end{equation}
and consider dissipation only through cavity losses
described by the term $\mathcal{D}[a]$  (see equation~\eqref{LindbladD}) but neglect spontaneous emission (e.g. described by $\mathcal{D}[J_-]$).
The Liouville operator is 
\begin{equation}
\mathcal{L} = \mathcal{L}_H(t) + \kappa \mathcal{D}[a] \;,
\end{equation}
with loss rate $\kappa \ge 0$.
In contrast to the standard quantum optical treatment in rotating wave approximation, it is not possible to eliminate the explicit time-dependence of $H(t)$ through a transformation to the rotating frame.

\begin{figure}
\includegraphics[width=0.46\textwidth]{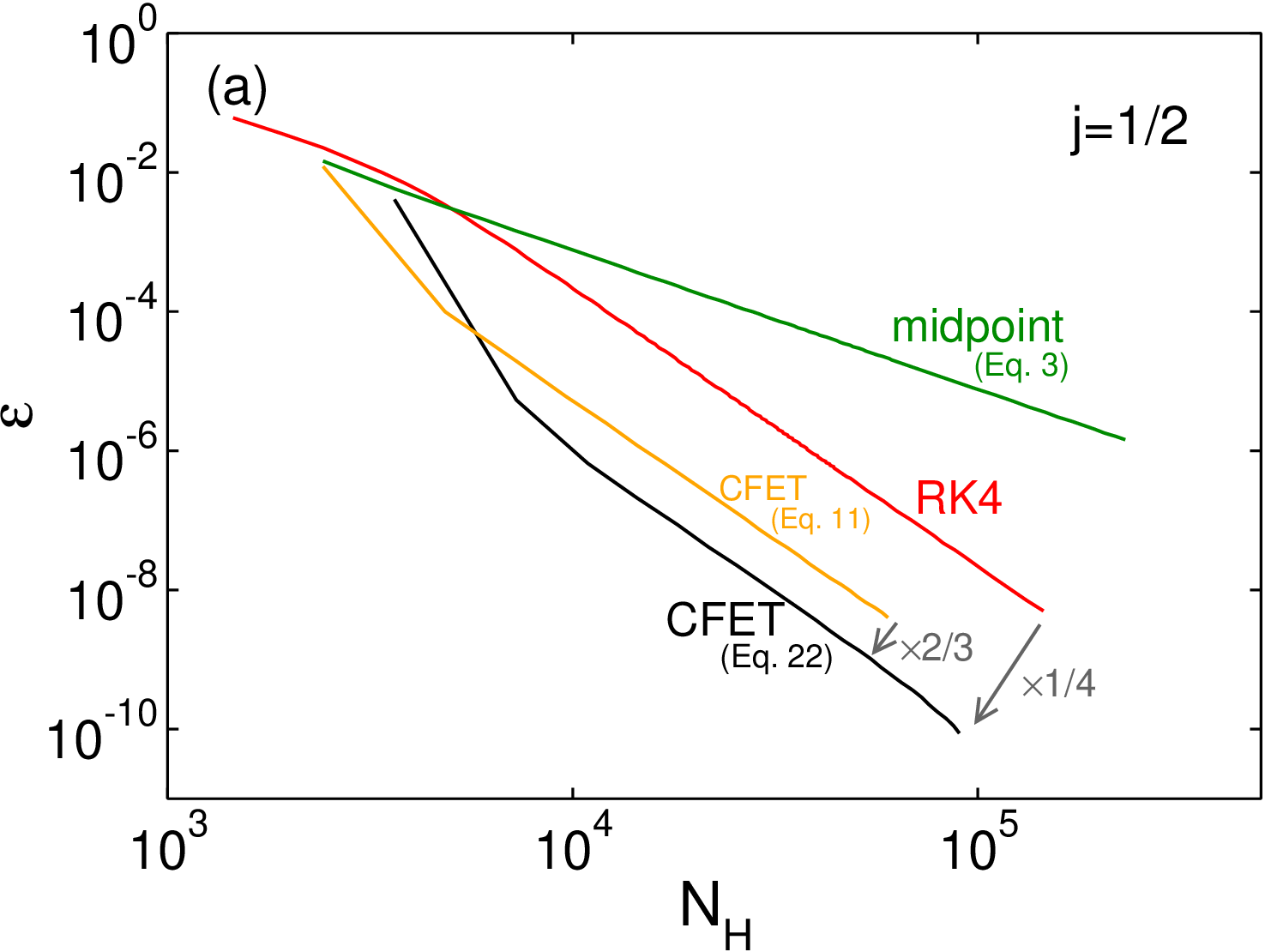} \hfill
\includegraphics[width=0.46\textwidth]{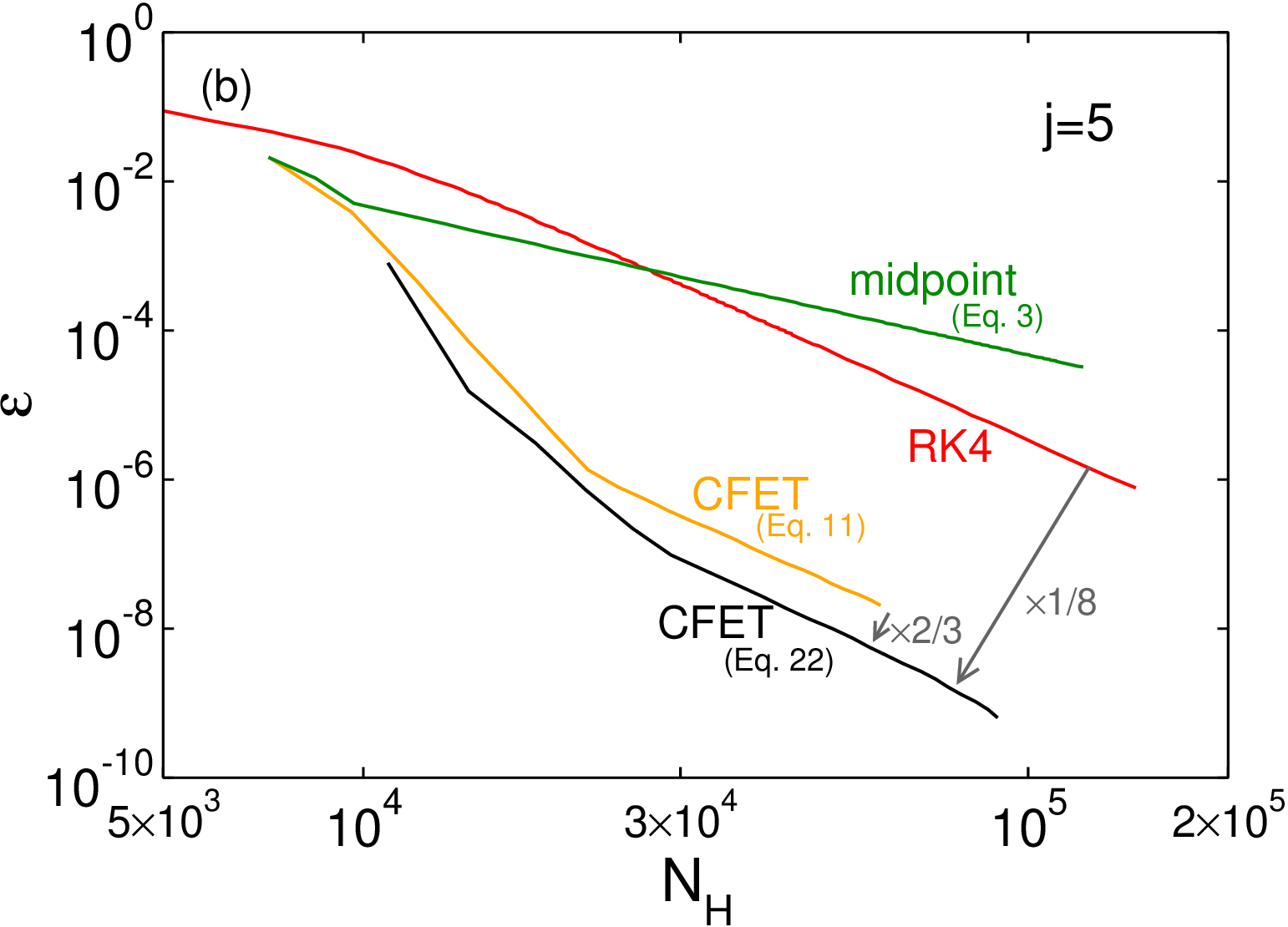} 
\caption{Comparison of the recommended (equation~\eref{CFET}) and simpler (equation~\eref{SimpleCFET}) CFET with the RK4 procedure and the middle-point approximation~\eref{Middle} in application to the parametrically driven dissipative Dicke model.
The Dicke model, with parameters $\Delta=\Omega=1$, $\kappa=0.01$, $\lambda_0 = 1$, $\delta \lambda=0.5$, $\omega_p=2$, is propagated over 30 modulation periods, i.e for $0 \le
t \le 30 (2\pi/\omega_p)$.
 As in figure~\ref{fig:Spin1} we show the error $\epsilon$ versus the effort $N_H$ 
for all propagation schemes.
The error is determined from comparison of the numerical density matrix and the reference solution obtained in the limit $N_H \to \infty$.
The CFETs are used in combination with a Chebyshev evaluation of the matrix exponential.
Left panel: For $j=1/2$, 
the CFET~\eref{CFET} is 4 times more efficient than the RK4 procedure.
Right panel: For $j=5$,
the CFET~\eref{CFET} is 8 times more efficient than the RK4 procedure.
} 
\label{fig:DickeCFET}
\end{figure}

Because the rotating wave approximation is not applicable,
physical properties of the parametrically driven dissipative Dicke model have to be extracted from time propagation of the density matrix of the joint atom-photon system.
The number of entries of the density matrix, which grows as $\propto (2j+1)^2$,
becomes large already for moderate pseudo-spin length $j$.
In addition, highly excited states contribute to the dynamics  if $j$ grows.
Therefore, as we discussed in section~\ref{sec:NumProb}, the advantage of CFETs over general ODE solvers such as the RK4 procedure will be pronounced for this more complex example.

In figure~\ref{fig:DickeCFET} we compare the recommended CFET from equation~\eref{CFET} to the RK4 procedure and the naive middle-point approximation~\eref{Middle}.
We see that already for $j=5$ we can easily reduce the numerical effort by a factor eight if we use CFETs. The reduction is achieved independently of the intended error $\epsilon$.
The middle-point approximation, which is only a second order scheme, is  not able to compete even with the RK4 procedure.
This clearly supports our recommendation for the CFET~\eref{CFET}, and extends the positive results from~\cite{AF11} to dissipative systems.
Note that the CFET~\eref{CFET} is better (by 50\%) than the simpler CFET~\eref{SimpleCFET} although it requires computation of three instead of two matrix exponentials.

Note again that the advantage of CFETs stems from the fact that the propagator is approximated as an product of exponentials of the Hamilton or Liouville operator. 
We know that this strategy is favourable for non-dissipative systems because it respects the unitary geometry of the Schr\"odinger equation~\cite{BCOR09,HLW06}, but apparently it works well also for density matrix propagation in driven dissipative systems.
Conversely, CFETs rely on good computation of matrix exponentials.
As mentioned at the beginning of section~\ref{sec:App} we use the Chebyshev technique here because it proved to be more efficient and reliable than a Krylov computation for non-hermitian matrices.

\subsection{Steady state resonances}

\begin{figure}
\includegraphics[width=0.55\textwidth]{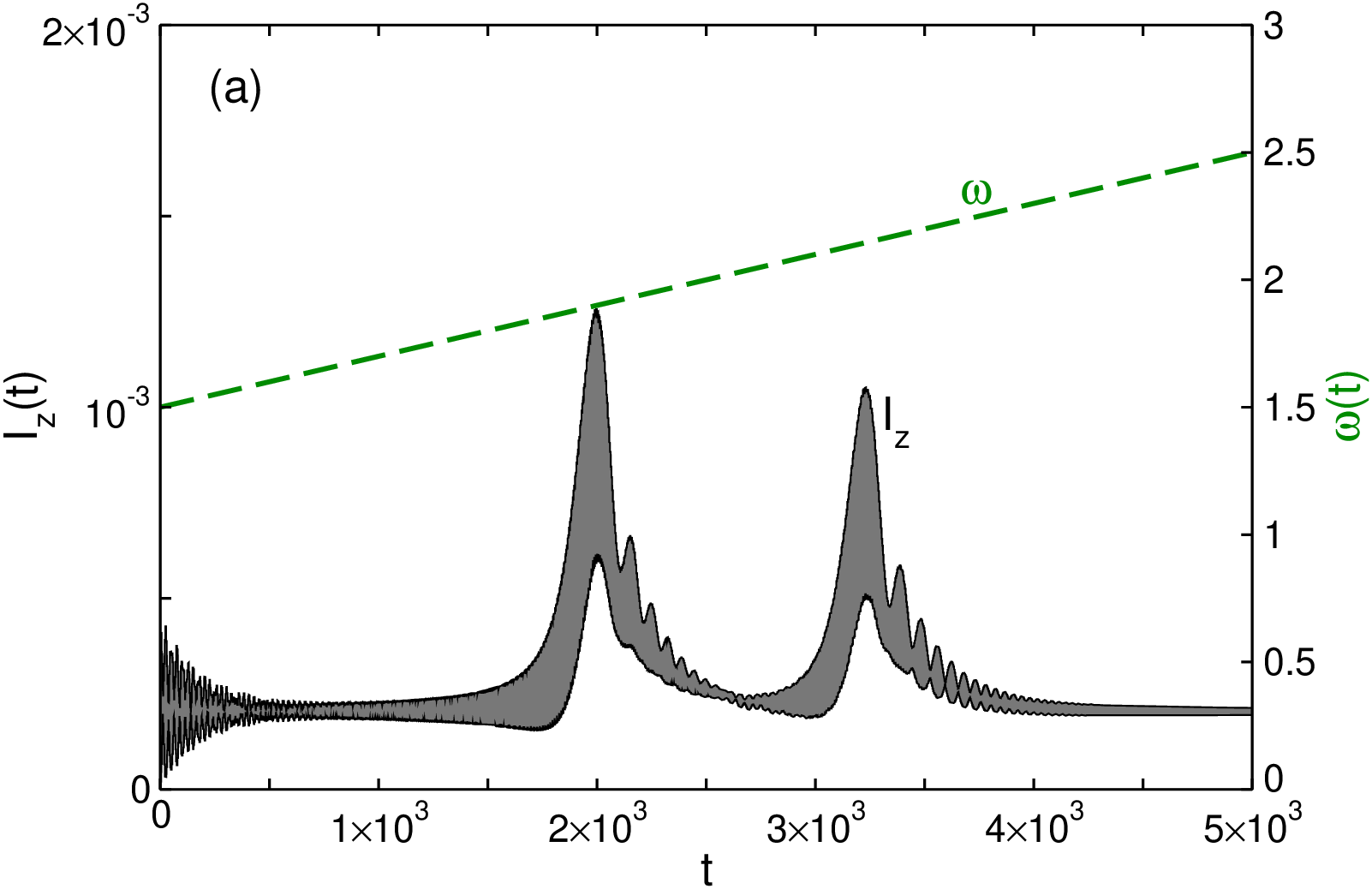}
\hfill
\includegraphics[width=0.4\textwidth]{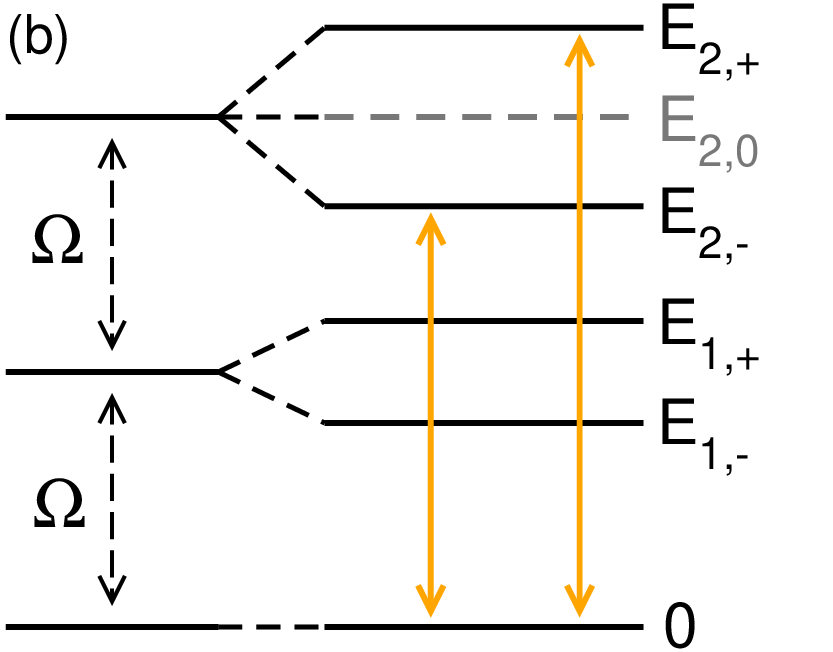}
 \caption{Panel (a): Time-dependence of $I_z(t)$ in the parametrically driven dissipative Dicke model for $j=5$, with a slow linear change of the modulation frequency $\omega_p$ from $1.5$ to $2.5$.
 The parameters are $\Omega=\Delta=1$, $\kappa= 5 \times 10^{-3}$, $\lambda_0=0.02$, and $\delta \lambda/\lambda_0 = 2.5 \times 10^{-2}$. 
 Shown is the envelope function of $I_z(t)$ instead of the fast oscillations with frequency $\Delta$.
 Panel (b): Lowest energy levels of the Dicke model at weak coupling and $\Omega=\Delta$.
  In the Rabi case  $j=1/2$ only two doubly excited states with energies $E_{2,\pm}$ exist,
 for $j>1/2$ an additional third state with energy $E_{2,0}$ appears.
 Transitions from the ground state change the number of excitations by two, and follow the vertical arrows.
 The matrix element for the transition with energy $E_{2,0}$ vanishes in lowest order perturbation theory.
 }
 \label{fig:Dicke1}
\end{figure}

In panel (a) in figure~\ref{fig:Dicke1} we show the time evolution of the initial state $\rho(0) = |\psi\rangle \langle \psi|$,
where $|\psi\rangle = |{-j/2}\rangle \otimes |\mathrm{vac}\rangle$ is the state with no atomic or field excitations.
We plot the population inversion
\begin{equation}
 I_z(t) = \frac{1}{2} + \frac{1}{j} \langle J_z (t) \rangle
\end{equation}
for a linear variation of the modulation frequency $\omega_p$ from $1.5$ to $2.5$.
Transient oscillations are observed for $t \lesssim 500$,
before two resonances evolve at a time corresponding to $\omega_p \approx 2 \pm 0.12$.
Beyond the resonances, $I_z(t)$ decays again to a small value with weak oscillations.

The energy level diagram in panel (b) in figure~\ref{fig:Dicke1} explains the appearance of resonances in panel (a) through transitions between the states of the Jaynes-Cummings ladder~\cite{SZ97}.

The eigenstates of the zero coupling Hamiltonian are the $J_z$, $a^\dagger a$ eigenstates $|m,n\rangle$, with $m+j$ atomic excitations and $n$ photons.
For $\Omega=\Delta$ the states 
$|{m+k,n-k}\rangle$, for several integer $k$,
are degenerate with energy $(m+n) \Omega$.
At weak coupling $\lambda_0 \ll \Omega, \Delta$,
degenerate states are split $\propto \lambda_0$ by the atom-field coupling. 
Counting energies relative to the energy of the lowest state $|{-j,0}\rangle$,
the energies of the two singly excited states $(|{-j+1,0} \rangle \pm |{-j,1} \rangle)/\sqrt{2}$ are given by 
$E_{1,\pm} = \Omega \pm \sqrt{2j} \lambda_0$.

A weak modulation of $\lambda(t)$ introduces transitions between $|{-j,0}\rangle$ and the doubly excited states (vertical arrows in panel (b)). 
Note that the splitting of energy levels in the diagram is determined by the co-rotating terms $J_- a^\dagger$, $J_+ a$ in the coupling term $J_x (a + a^\dagger)$, which preserve the number of excitations in the sense of the rotating wave approximation, but the transitions arise from the counter-rotating terms $J_+ a^\dagger$, $J_- a$ and change the number of excitations by two.
In the Rabi case $j=1/2$, the two doubly excited states $(|{1/2,1} \rangle \pm |{-1/2,2} \rangle)/\sqrt{2}$ have energy $E_{2,\pm} = 2\Omega \pm \sqrt{2} \lambda_0$.
Resonances are expected at these energies~\cite{LGCC09}.

In the Dicke case $j>1/2$, the splitting of the three degenerate states 
$|{-j+2,0}\rangle$,
$|{-j+1,1}\rangle$,
$|{-j,2}\rangle$
has to be calculated.
This gives the energies $E_{2,\pm} = 2 \Omega \pm  \sqrt{8j-2} \lambda_0$
for the odd/even parity combination, reproducing the $j=1/2$ result.
The third state with unshifted energy $E_{2,0} = 2\Omega$ is a linear combination
of $|{-j+2,0}\rangle$, $|{-j,2}\rangle$ only.
It does not couple to $|{-j,0}\rangle$ through the counter-rotating terms,
and the transition to this state is forbidden in leading order of perturbation theory.
Therefore, we also expect only two resonances in the Dicke case.
For the parameters from figure~\ref{fig:Dicke1}, they occur at $E_{2,\pm} = 2 \pm 0.02 \sqrt{38} \approx 2 \pm 0.1233$.

\begin{figure}
\includegraphics[width=0.48\textwidth]{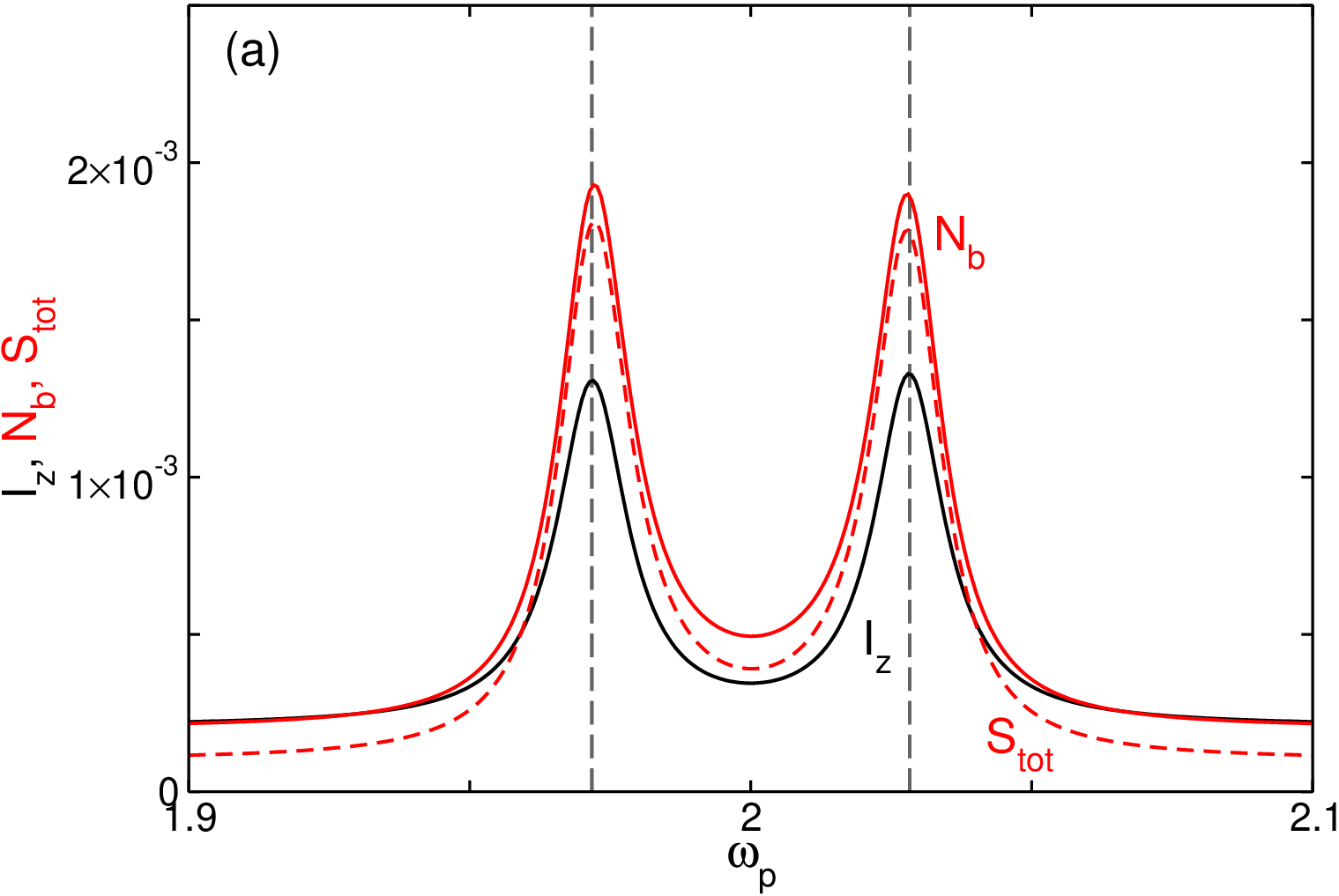} \hfill
\includegraphics[width=0.48\textwidth]{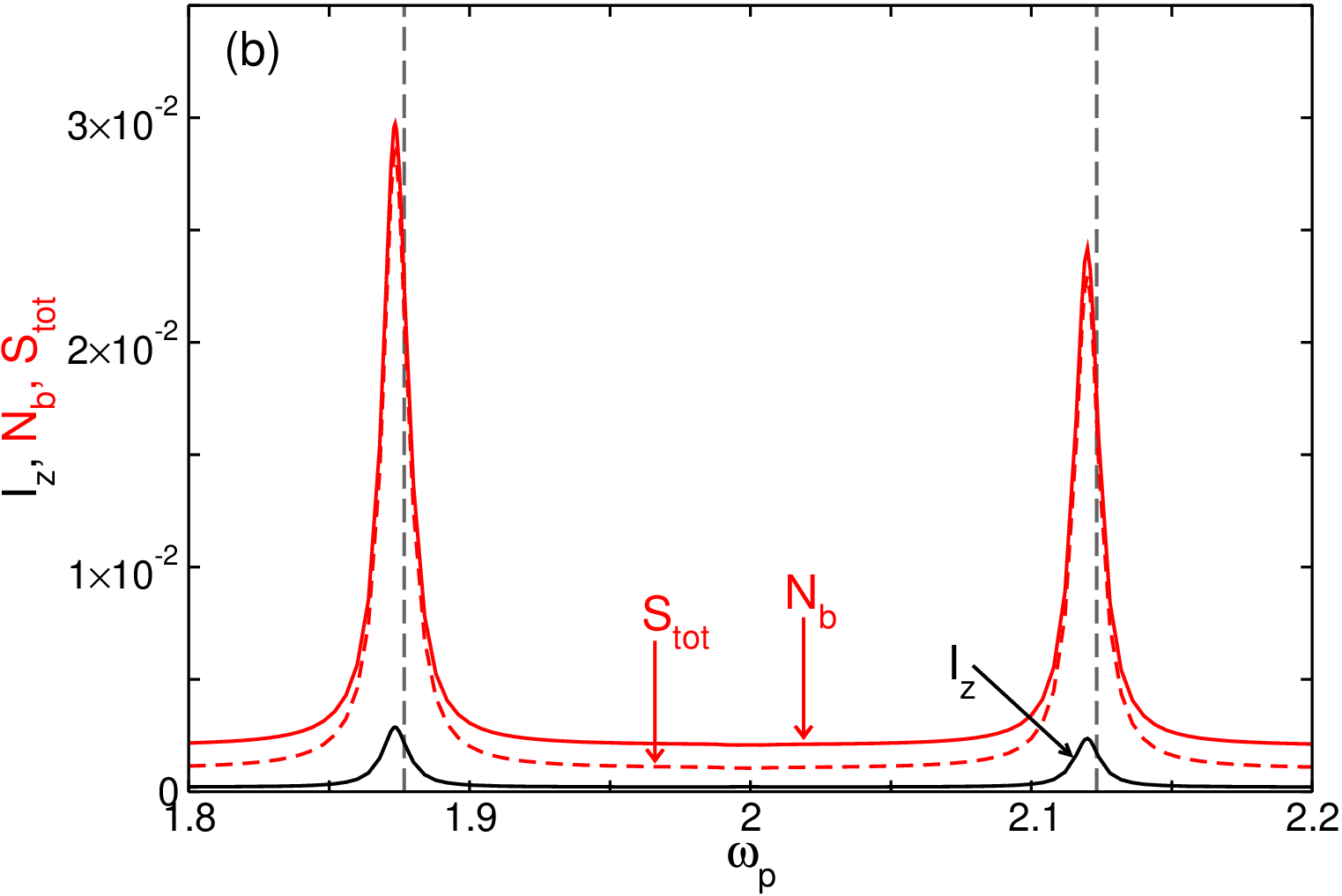} 
 \caption{Population inversion $I_z$ (black curve), number of cavity bosons $N_b$ (red curve), and total emission $S_\mathrm{tot}$ (dashed red curve) for the driven dissipative Dicke model, 
 as a function of the modulation frequency $\omega_p$
 with parameters $\Omega=\Delta=1$, $\kappa=5 \times 10^{-3}$, $\lambda_0=0.02$,  $\delta \lambda/\lambda_0 = 2.5 \times 10^{-2}$ as in figure~\ref{fig:Dicke1}.
 The quantities are averaged over one modulation period $2\pi/\omega_p$.
 Panel (a): Results for the Rabi case $j=1/2$.
 Panel (b): Results for $j=5$. 
 The grey dashed lines indicates the resonances at  $\omega_p = E_{2,\pm}$.
 }
 \label{fig:Dicke2}
\end{figure}

To identify the resonances numerically
we propagate the system with fixed $\omega_p$ until the periodic steady state is reached.
Then we calculate the quantities 
\begin{equation} 
I_z = \int_{t}^{t+ 2\pi/\omega_p} I_z(t') \, dt' \;, 
\quad N_b  = \int_{t}^{t+ 2\pi/\omega_p} N_b(t') \, dt' 
\end{equation}
 averaged over one modulation period $2 \pi/\omega_p$.
 Here,
 \begin{equation}
  N_b(t) = \langle a^\dagger(t) a(t) \rangle \;.
\end{equation}
is the number of cavity bosons.

In figure~\ref{fig:Dicke2} the quantities $I_z$, $N_b$ are shown as a function of $\omega_p$.
We recognize the two resonances $\omega_p \approx E_{2,\pm}$,
which are broadened due to the cavity losses $\propto \kappa$.
For the calculation we used the optimized 4th-order CFET~\eqref{CFET} together with a Chebyshev computation of the exponential.

\subsection{Emission spectrum}

To study the optical properties of this system we compute the cavity emission spectrum $S(\omega)$.
It is obtained as the Fourier transform
\begin{equation}
  S(\omega) = \frac{1}{\pi} \, \Re \int_0^\infty S(\tau) e^{-\ii \omega t} d \tau
\end{equation}
of the correlation function
\begin{equation}
 S(\tau) = \int_{t}^{t+ 2\pi/\omega_p} \, \langle a^\dagger (t'+\tau) a(t') \rangle \, dt' \;,
 \end{equation}
 which we calculate with the quantum regression theorem~\cite{Carm99} through time propagation of the operator $a \rho(t)$ (for $\tau \ge 0$).
The correlation function involves the average over one modulation period $[t,t + 2\pi/\omega_p]$ for large $t$, i.e. in the periodic steady state.

We include in figure~\ref{fig:Dicke2} the total emission
\begin{equation}
 S_\mathrm{tot} = \int_0^\infty S(\omega) d\omega \;,
\end{equation}
which is given by the integral over positive $\omega$ in accordance with the fact that emission of a (real) photon can only decrease the energy.
We note the normalization
\begin{equation}
 \int_{-\infty}^\infty S(\omega) d\omega = S(\tau=0) = N_b  \;.
\end{equation}
We see that $S_\mathrm{tot} \approx N_b$ close to resonance, when emission is strong.
Away from resonance $S_\mathrm{tot}$ drops below $N_b$,
since $N_b$ counts also bound photons that do not contribute to emission.
However, $S_\mathrm{tot}$ remains finite as a consequence of the Markovian approximation used here for the dissipative term~\cite{LGCC09}.

\begin{figure}
\includegraphics[width=0.38\textwidth]{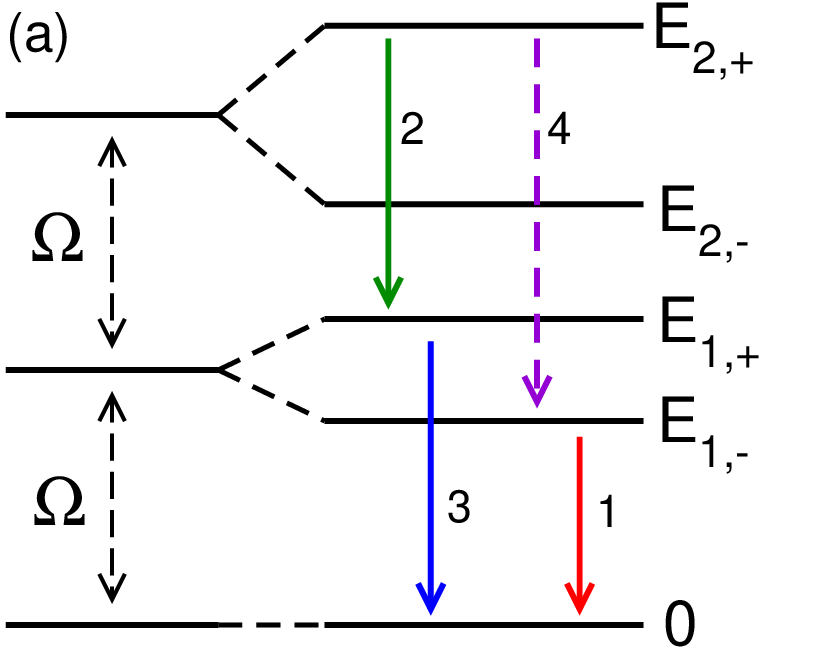} \hfill
\includegraphics[width=0.6\textwidth]{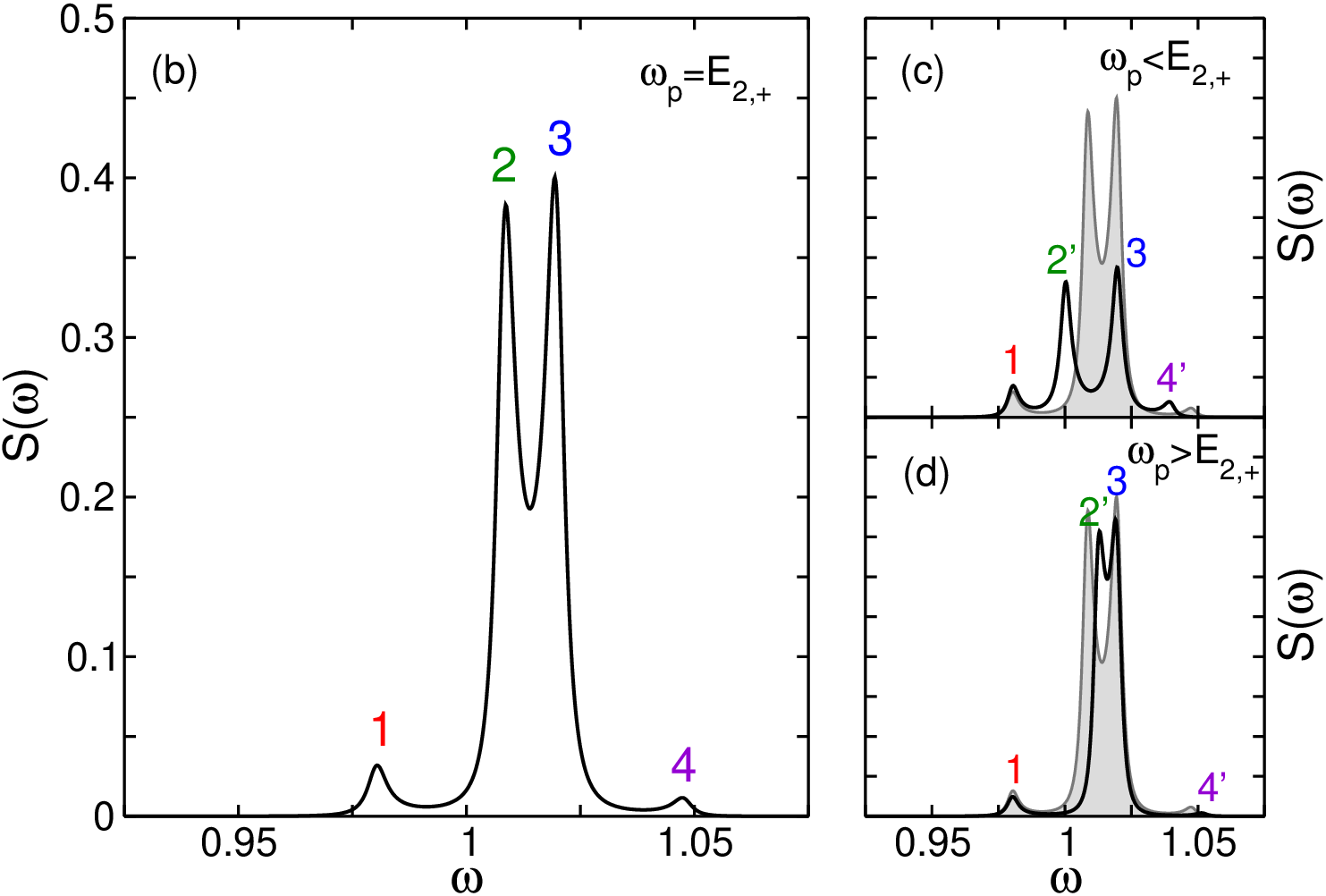} 
 \caption{Emission spectrum $S(\omega)$ for the Rabi case $j=1/2$,
 at (panel (b)) and close to (panels (c), (d)) the higher resonance $\omega_p = E_{2,+}$.
 It is $\omega_p=E_{2,+}-0.008 \Omega$ in panel (c) and
 $\omega_p=E_{2,+}+0.004 \Omega$ in panel (d).
 The remaining parameters are as in figure~\ref{fig:Dicke2}.
 The energy level diagram in panel (a) follows figure~\ref{fig:Dicke1}.
 The four transitions are marked by vertical arrows and indicated by corresponding numbers in the other panels. The transition energies are given in equation~\eqref{TransErgRabi1}.
 The spectrum from panel (b) is included in panels (c), (d) as the grey filled curve.
 }
 \label{fig:RabiSpec1}
\end{figure}

\begin{figure}
\includegraphics[width=0.38\textwidth]{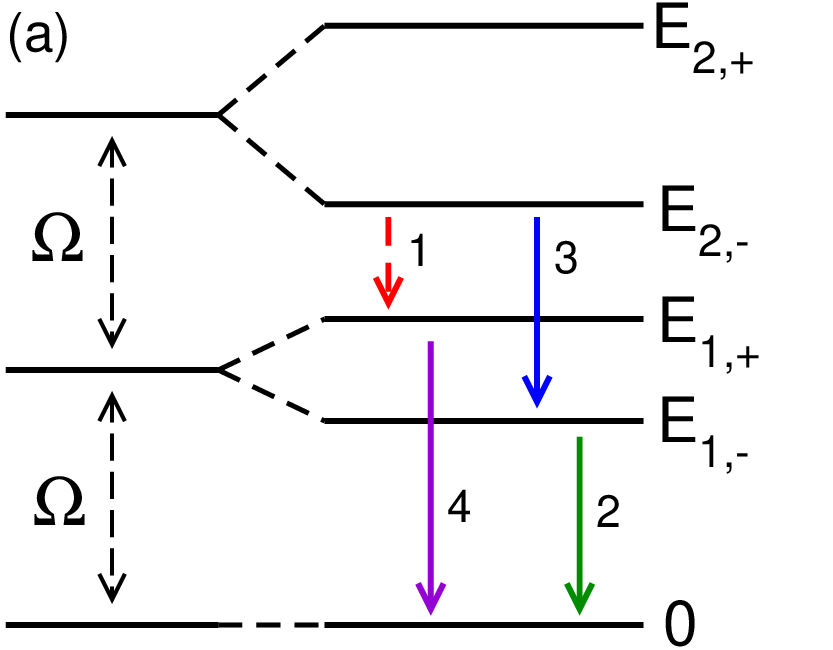} \hfill
\includegraphics[width=0.6\textwidth]{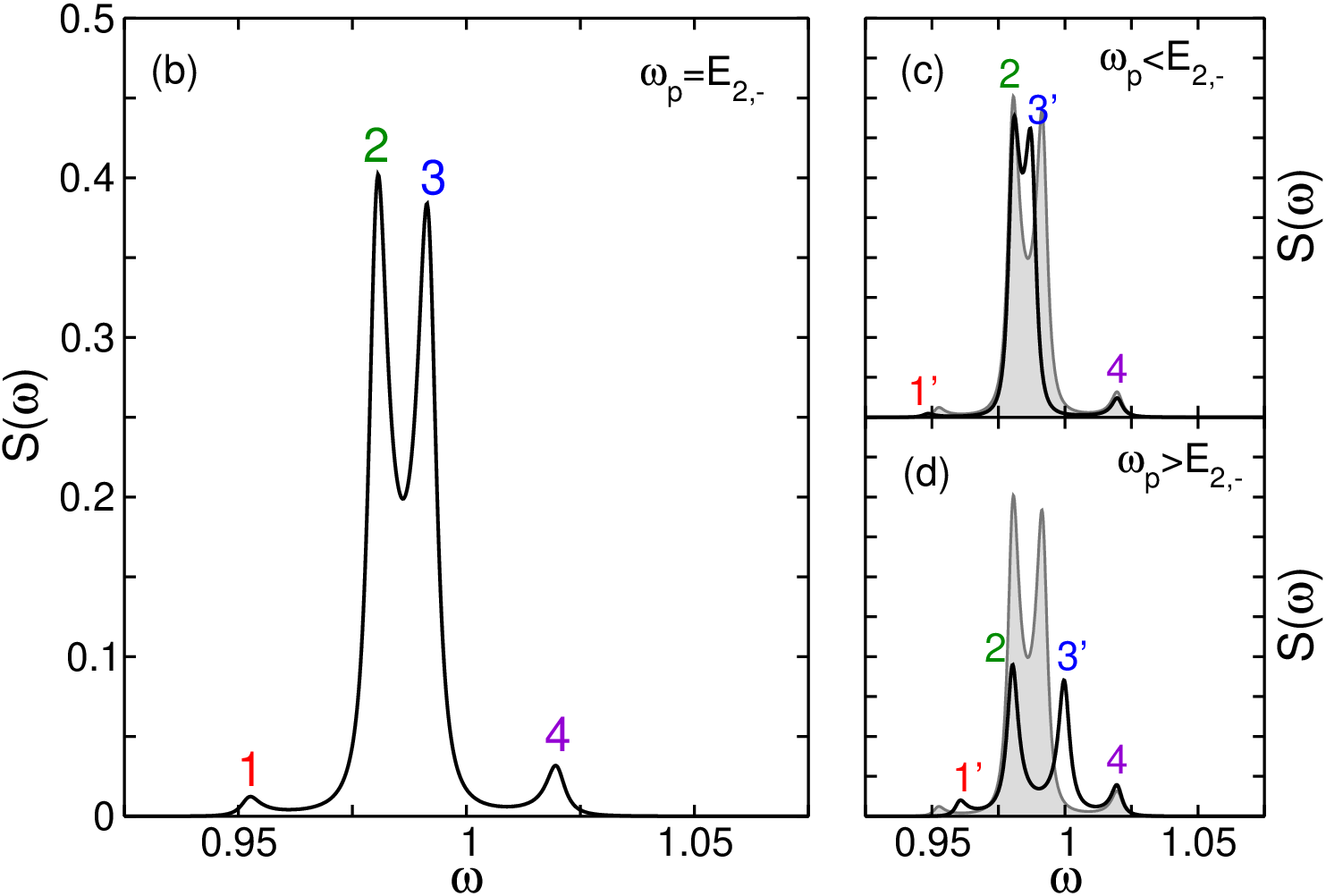} 
 \caption{ Emission spectrum $S(\omega)$ for the Rabi case $j=1/2$,
 at (panel (b)) and close to (panels (c), (d)) the lower resonance $\omega_p = E_{2,-}$.
 It is $\omega_p=E_{2,-}-0.004 \Omega$ in panel (c) and
 $\omega_p=E_{2,-}+0.008 \Omega$ in panel (d).
 The remaining parameters are as in figure~\ref{fig:Dicke2},
 and the notation follows the previous figure~\ref{fig:RabiSpec1}.
 }
 \label{fig:RabiSpec2}
\end{figure}

\begin{figure}
\includegraphics[width=0.38\textwidth]{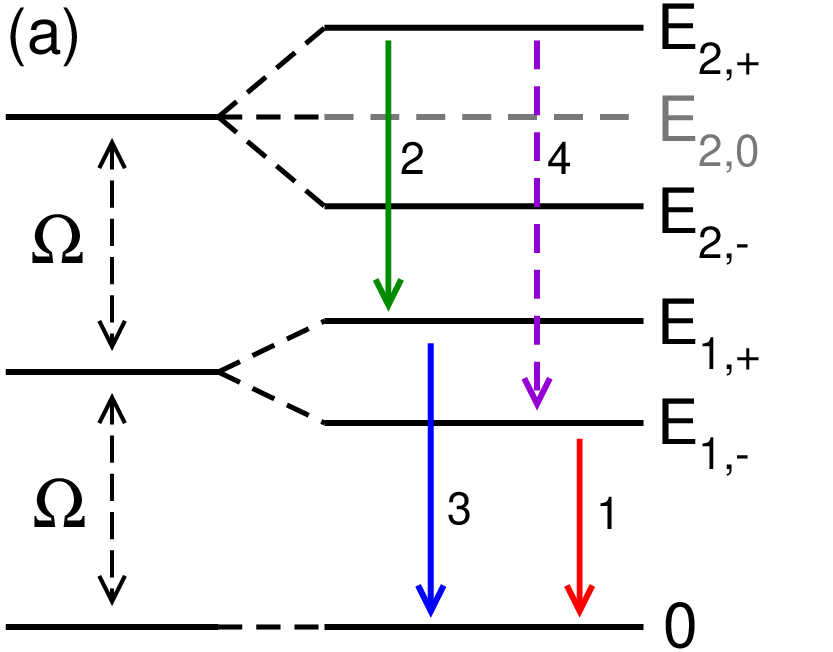} \hfill
\includegraphics[width=0.6\textwidth]{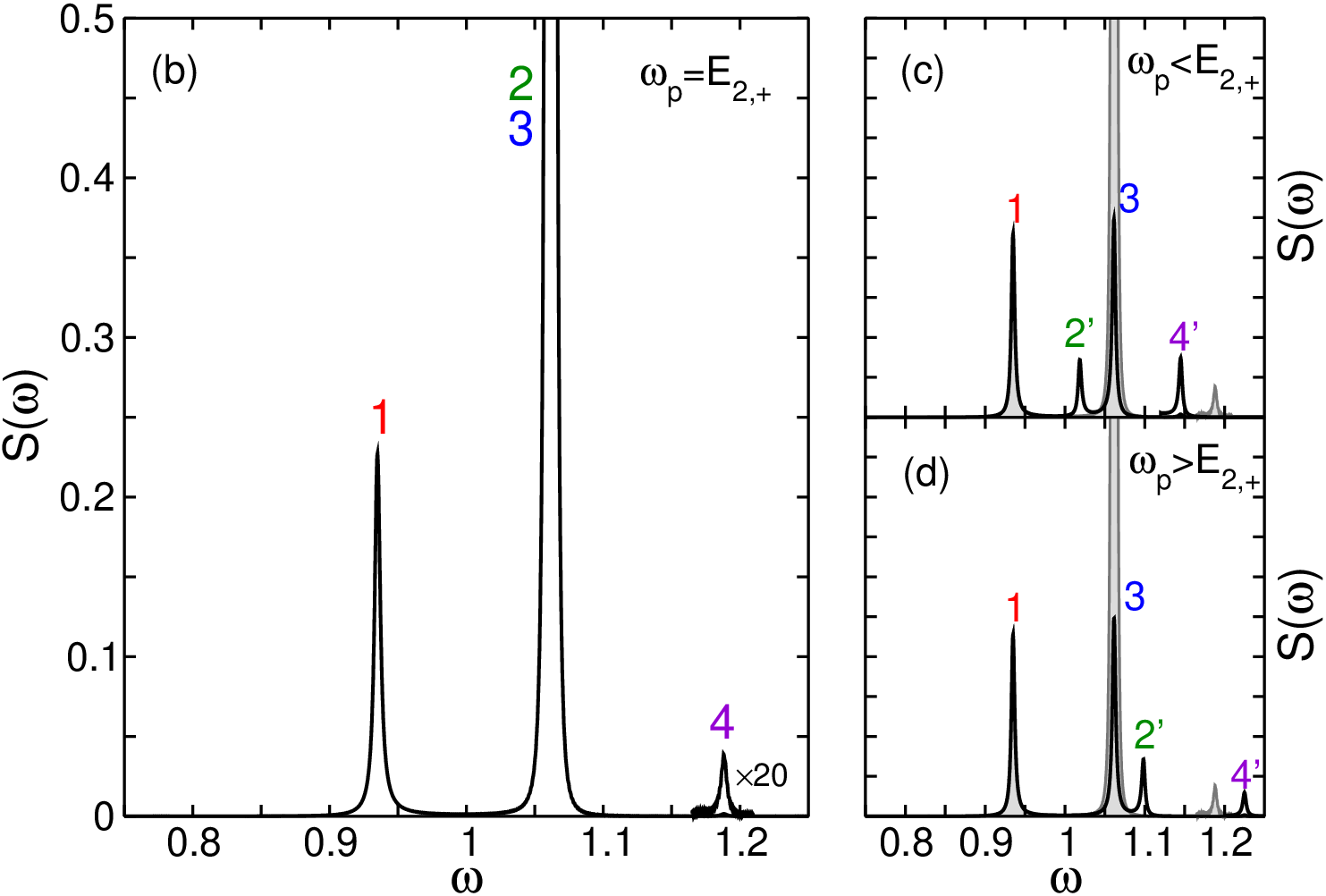} 
 \caption{
 Emission spectrum $S(\omega)$ for $j=5$,
 at (panel (b)) and close to (panels (c), (d)) the higher resonance $\omega_p = E_{2,+}$.
 It is $\omega_p=E_{2,+}-0.04 \Omega$ in panel (c) and
 $\omega_p=E_{2,+}+0.04 \Omega$ in panel (d).
 The remaining parameters are as in figure~\ref{fig:Dicke2},
 and the notation follows the previous figures~\ref{fig:RabiSpec1},~\ref{fig:RabiSpec2}. 
 The transition energies are given in equation~\eqref{TransErgDicke1}.
 The weak signals of the parity-forbidden transitions $4$, $4'$ are displayed with magnification factor $20$ in all panels.
 }
 \label{fig:DickeSpec1}
\end{figure}

\begin{figure}
\includegraphics[width=0.38\textwidth]{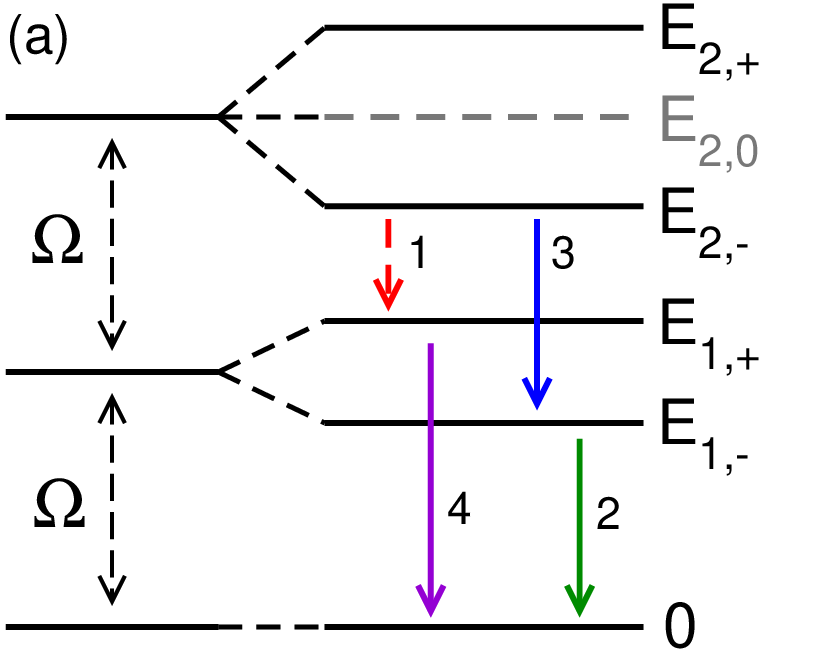} \hfill
\includegraphics[width=0.6\textwidth]{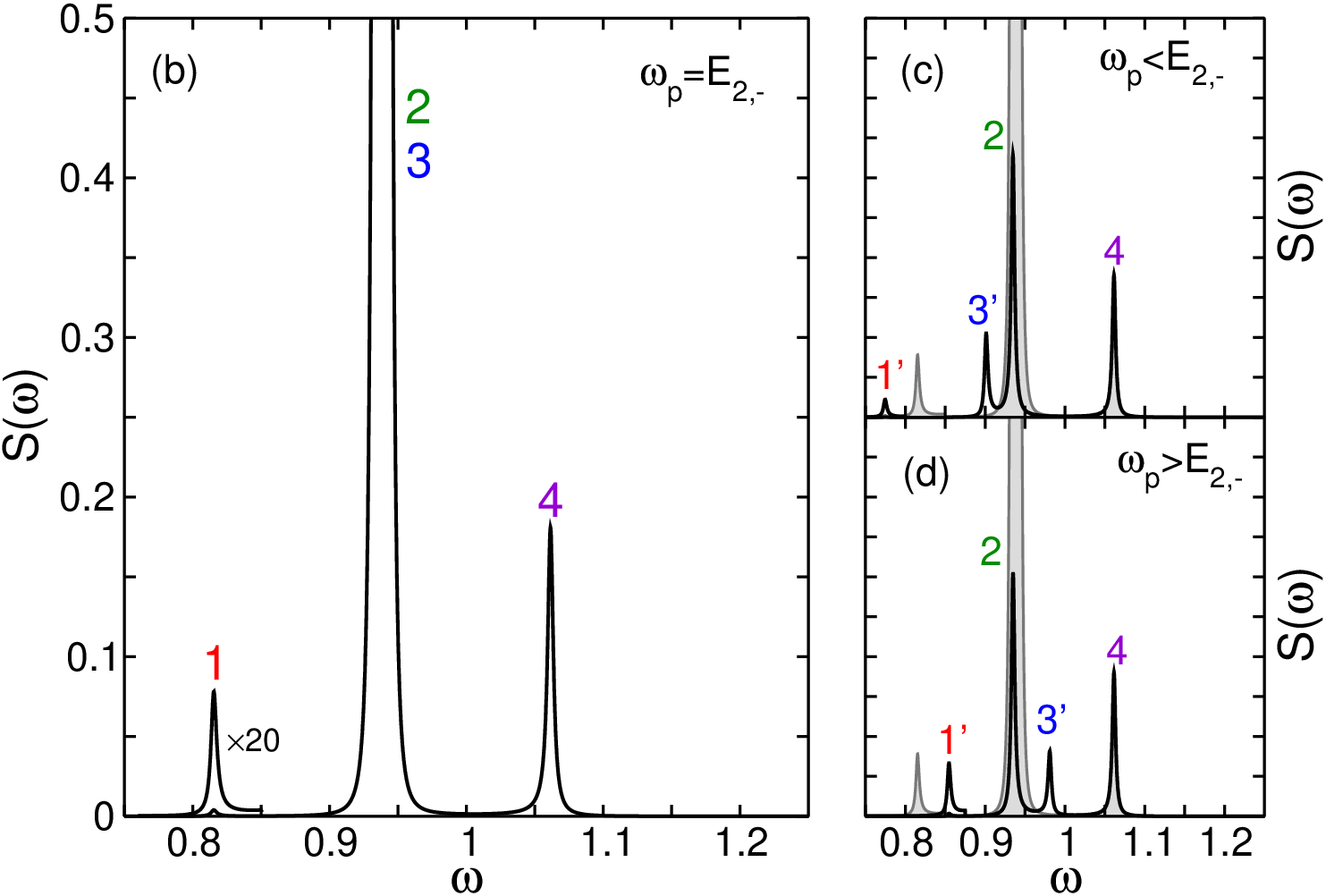} 
 \caption{
 Emission spectrum $S(\omega)$ for $j=5$,
 at (panel (b)) and close to (panels (c), (d)) the lower resonance $\omega_p = E_{2,-}$.
 It is $\omega_p=E_{2,-}-0.04 \Omega$ in panel (c) and
 $\omega_p=E_{2,-}+0.04 \Omega$ in panel (d).
 The remaining parameters are as in figure~\ref{fig:Dicke2},
 and the notation follows the previous figures~\ref{fig:RabiSpec1}--\ref{fig:DickeSpec1}.
 The weak signals of the parity-forbidden transitions $1$, $1'$ are displayed with magnification factor $20$ in all panels.
 }
 \label{fig:DickeSpec2}
\end{figure}

The emission spectrum for the Rabi case $j=1/2$ is shown in figures~\ref{fig:RabiSpec1},~\ref{fig:RabiSpec2} and for the Dicke case with $j=5$ in figures~\ref{fig:DickeSpec1},~\ref{fig:DickeSpec2}.
For weak coupling, i.e. for $|\lambda(t)| \ll \{\Delta, \Omega\} $, the interpretation of the emission spectrum is again possible using the energy level diagram from figure~\ref{fig:Dicke1}.

For $\omega_p \approx E_{2,+}$ in figure~\ref{fig:RabiSpec1}, the 
higher of the two doubly excited states is populated.
Since the operator $a$ changes the number of excitations by one,
an atom in this state can decay to the lowest state only through the intermediate singly excited states.
Four transitions corresponding to the four vertical arrows in panel (a) can be identified in this situation.
For the Rabi case $j=1/2$,
they are in order of increasing energy
\begin{eqnarray}\label{TransErgRabi1}
\nonumber \fl\quad \omega_1 &= E_{1,-} = \Omega - \lambda_0 \approx 0.98 \;, \quad
 \omega_2 &= E_{2,+}-E_{1,+} = \Omega + \lambda_0 (\sqrt{2}-1) \approx  1.01 \;, \quad \\
 \fl\quad \omega_3 &= E_{1,+} = \Omega + \lambda_0 \approx  1.02 \;, \quad
 \omega_4 &= E_{2,+} - E_{1,-} = \Omega + \lambda_0 (\sqrt{2}+1) \approx  1.05 \;.
\end{eqnarray}
The numerical values correspond to the parameters from figure~\ref{fig:RabiSpec1},
and the transitions are marked correspondingly in panels (a), (b).

In transitions $2$ and $4$ the doubly excited state decays.
These transitions are shifted by $\omega_p - E_{2,+}$ away from resonance
(transitions $2'$, $4'$ in panels (c) and (d)).
In transitions $1$ and $3$ the singly excited states decay,
and their energy does not depend on the modulation frequency.
Note that transition $4$ (dashed arrow) is parity forbidden in lowest order perturbation theory and gives only a weak signal.

The opposite case $\omega_p \approx E_{2,-}$ in figure~\ref{fig:RabiSpec2}
has an analogous interpretation that follows from the energy diagram in panel (a).
Now the lower of the doubly excited states is populated,
and the order of transitions is reversed, with transition $1$ as the parity forbidden transition.

\begin{figure}
\includegraphics[width=0.46\textwidth]{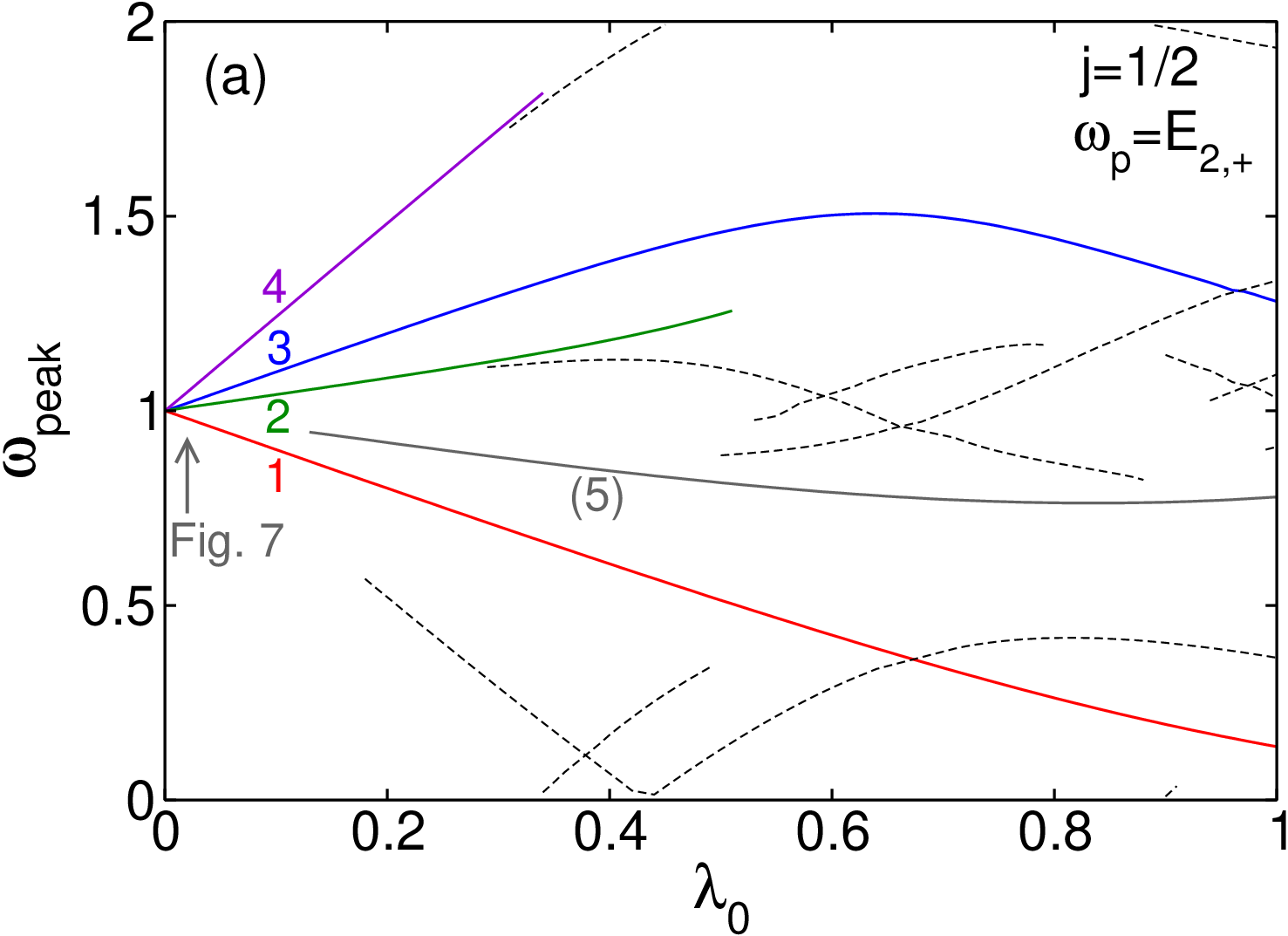} \hfill
\includegraphics[width=0.46\textwidth]{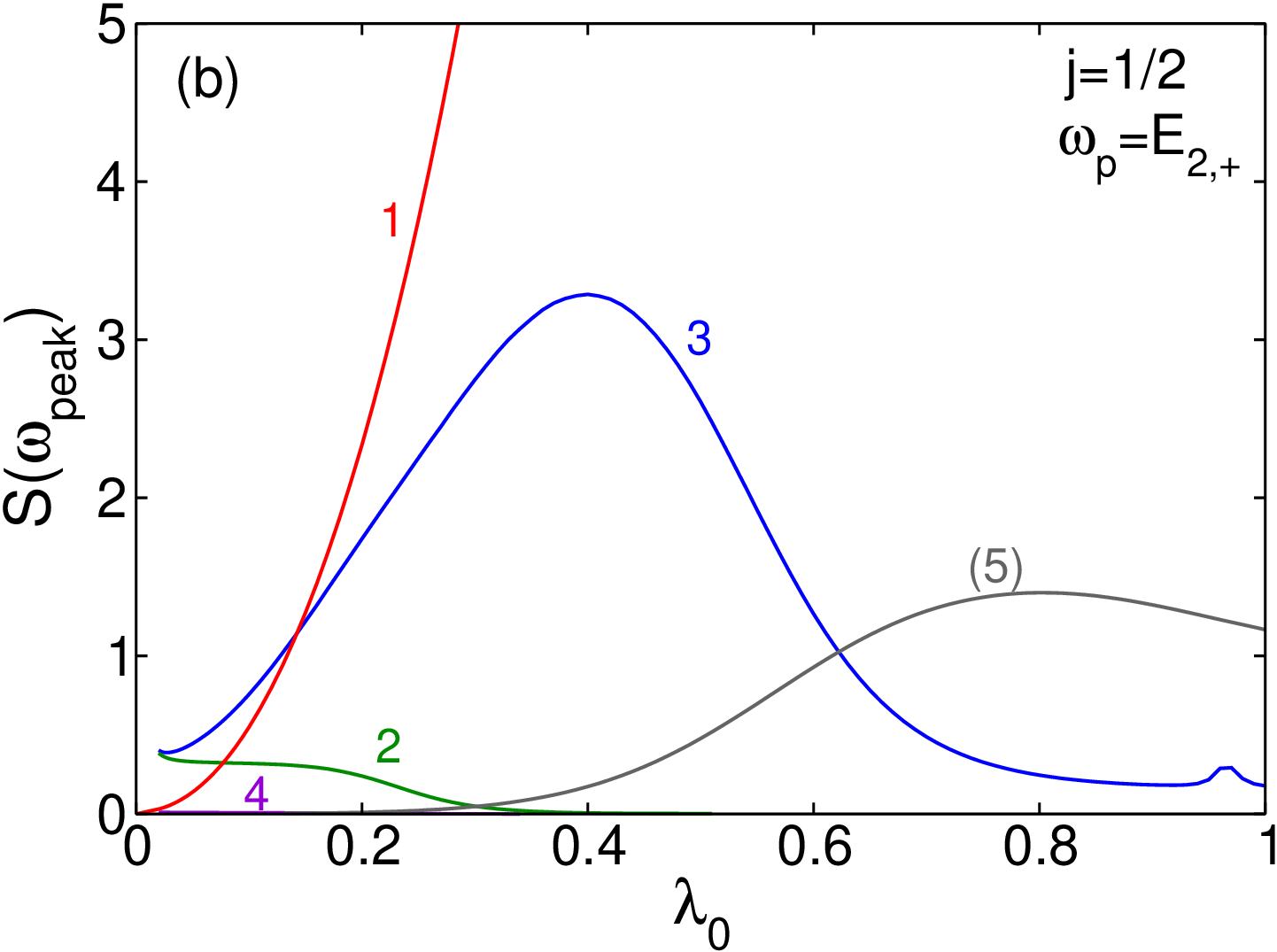} 
\caption{Peaks in the emission spectrum $S(\omega)$ at stronger coupling, for the Rabi case $j=1/2$ at the higher resonance $\omega_p=E_{2,+}$. Left panel: Energies $\omega_\mathrm{peak}$ of the different peaks as a function of coupling $\lambda_0$. The solid curves marked 1--4 correspond to the transitions in figure~\ref{fig:RabiSpec1} at $\lambda_0=0.02$, the dashed curves show the position of additional peaks emerging at stronger coupling.
Because the peaks in $S(\omega)$ have a finite width, curves in the panel can begin and end in an isolated point.
Right panel: Peak height $S(\omega_\mathrm{peak})$ of the five highlighted curves in the left panel. The height of the parity forbidden transition $4$ (cf. Panel (a) in figure~\ref{fig:RabiSpec1}) is close to zero for all $\lambda_0$.
}
\label{fig:RabiStrong}
\end{figure}

For the Dicke case $j > 1/2$ in figures~\ref{fig:DickeSpec1},~\ref{fig:DickeSpec2} we expect the same qualitative behavior as for $j=1/2$ since the additional state with energy $E_{2,0}$ does not participate in the transitions.
For the higher resonance $\omega_p \approx E_{2,+}$ in figure~\ref{fig:DickeSpec1},
the situation corresponding to figure~\ref{fig:RabiSpec1}, the transitions in order of increasing energy are
\begin{eqnarray}\label{TransErgDicke1}
\fl \nonumber \quad\omega_1 & = \Omega - \lambda_0 \sqrt{2j} \approx 0.94 \;, \quad
 \omega_2  &= \Omega + \lambda_0 (\sqrt{8j-2}-\sqrt{2j})\approx 1.06\;, \quad \\
\fl \quad \omega_3 &=  \Omega + \lambda_0 \sqrt{2j}\approx 1.06 \;, \quad
 \omega_4 &=  \Omega + \lambda_0 (\sqrt{8j-2}+\sqrt{2j})\approx 1.19\;.
\end{eqnarray}
The numerical values correspond to the parameters from figure~\ref{fig:DickeSpec1}.

As a difference to the Rabi case $j=1/2$
we note that $\omega_2 \approx \omega_3$, 
such that the two transitions cannot be distinguished in panel (b) because of the finite linewidth acquired through cavity losses.
If we change $\omega_p$ transition $2$ is shifted but transition $3$ remains fixed. This allows for the separate identification of both peaks in panels (c), (d).
In the same way we can understand the opposite case $\omega_p \approx E_{2,-}$ in figure~\ref{fig:DickeSpec2}, which is similar to figure~\ref{fig:RabiSpec2}.

For stronger coupling, additional peaks appear in the emission spectrum.
The peak position and height is shown in Fig.~\ref{fig:RabiStrong} for the situation corresponding to figure~\ref{fig:RabiSpec1}. Note that we adjust the modulation frequency $\omega_p=E_{2,+}$ for different $\lambda_0$ to remain close to the higher resonance. 
For $\lambda_0 \gtrsim 0.1$ a fifth transition peak ``(5)'' becomes visible, and the previous interpretation of $S(\omega)$ via the Jaynes-Cummings ladder breaks down. 
At least such situations require numerical time propagation because a simple perturbative interpretation is no longer possible.

\section{Summary and Outlook}\label{sec:Summary}

Numerical time propagation allows for the theoretical description of experimentally relevant non-equilibrium situations beyond the linear response regime.
Such situations arise in particular if small systems such as atoms or molecules
are manipulated by strong radiation fields.
Important directions of research include the optical properties of ensembles showing collective behavior, such as polariton or exciton condensates~\cite{LEKMSS04,EL06}.
A characteristic optical signature,
as of the spatial shape and energy distribution of the optical emission~\cite{SVG94,SS10} 
or the coherence properties of the emitted light,
can provide the proof of existence for a condensate.
A fundamental theory of optical properties of collective phases of (quasi-) particles with finite lifetime requires the description of the driven open quantum system that is realized, e.g., by the excitonic condensate in a semiconductor~\cite{Sno06}.
Different but on the fundamental technical level related questions arise in the field of non-equilibrium transport problems~\cite{HJ08}.

Increasing complexity of the physical situation under study coincides with an increase of the computational effort, which underlines the need for powerful numerical algorithms.
We argued here in favor of commutator-free exponential time-propagators as a convenient alternative to the original Magnus series.
Shifting the focus of previous studies,
where CFETs were shown to be well suited for the time propagation of driven systems without dissipation, we here applied CFETs to open quantum systems.
Using the parametrically driven Dicke model as an illustrative example
we calculated the optical emission spectrum with this technique.

Conceptually, CFETs are recipes for the reduction of the original problem --- the solution of the Schr\"odinger or master equations with a time-dependent Hamilton operator ---
to the computation of matrix exponentials.
Because they replace the naive approximation of the time propagator by a single exponential per time step, as it is encoded in the second order middle-point approximation~\eref{Middle}, with a more sophisticated combination of exponentials, higher-order CFETs significantly reduce the numerical effort.
The particular appeal of CFETs is that they can be combined with any technique for the computation of matrix exponentials, for example the powerful Krylov (Arnoldi) or Chebyshev techniques in the context of large sparse matrix computations.
CFETs do not compete with such techniques, but instead serve the complementary purpose of achieving a favorable error-effort scaling also for equations with a time-dependent $H(t)$.
Therefore, CFETs should be of interest to anyone presently using Krylov (Arnoldi) or Chebyshev techniques in studies of driven quantum systems: It is straightforward and simple enough to add the CFET computational scheme from equations~\eref{CFET},~\eref{CFET2} to an existing program or implementation~\cite{Sid98,GNZBS09}.

In addition to the principal research directions mentioned above
many open problems arise within the more restricted context of the present work.
A notorious problem is the efficient evaluation of the matrix exponential for non-symmetric large sparse matrices, which is essentially a problem of polynomial approximation in the complex plane without precise knowledge of the approximation domain.
One may also question the principal usage of CFETs for dissipative systems,
where dynamical semi-groups replace the Lie group setting.
Modifications of CFETs can be tailored to this situation
and circumvent the negative time-step problem that occurs for methods beyond the 4th-order CFETs to which we restricted our present considerations.

A more physical question concerns the use of the Lindblad formalism in the description of optical emission.
The Markovian approximation is not entirely satisfactory here,
since it cannot distinguish between energy increasing (``virtual'') and energy decreasing (``real'') transitions in the emission process. For this reason, the total emission rate remains finite (of the order of the cavity loss rate $\kappa$) even without external pumping although it should drop to zero.
The use of non-Markovian master equations~\cite{LGCC09,BP02} might overcome these problems, which should also be relevant if we ask for the coherence properties of the emitted light~\cite{Carm99,VW06}. 
Another possibility is the combination of CFETs with polynomial techniques for the numerical representation of open quantum systems~\cite{AF08,AF09,CRHP10}.

We hope to be able to return to some of these issues soon,
which we had to leave unresolved here.
Presently, we can conclude that CFETs are one promising contribution to numerical time propagation of complex non-equilibrium quantum systems  and warrant further exploration.

\ack
 This work was supported by Deutsche Forschungsgemeinschaft via AL1317/1-2 and SFB 652 (project B5). Work at Argonne was supported by DOE-BES under FWP70069.

\bigskip

\providecommand{\newblock}{}


\begin{thebibliography}{10}
\expandafter\ifx\csname url\endcsname\relax
  \def\url#1{{\tt #1}}\fi
\expandafter\ifx\csname urlprefix\endcsname\relax\def\urlprefix{URL }\fi
\providecommand{\eprint}[2][]{\url{#2}}

\bibitem{WVEB06}
Walther H, Varcoe B~T~H, Englert B~G and Becker T 2006 {\em Rep. Prog. Phys.\/}
  {\bf 69} 1325

\bibitem{BP02}
Breuer H~P and Petruccione F 2002 {\em The Theory of Open Quantum Systems\/}
  (Oxford University Press)

\bibitem{Carm99}
Carmichael H~J 1999 {\em Statistical Methods in Quantum Optics 1\/} (Springer)

\bibitem{VW06}
Vogel W and Welsch D~G 2006 {\em Quantum Optics: An Introduction\/} (Wiley-VCH)

\bibitem{Ma54}
Magnus W 1954 {\em Comm. Pure Appl. Math.\/} {\bf VII} 649

\bibitem{BCOR09}
Blanes S, Casas F, Oteo J~A and Ros J 2009 {\em Physics Reports\/} {\bf 470}
  151

\bibitem{BM06}
Blanes S and Moan P~C 2006 {\em App. Num. Math.\/} {\bf 56} 1519

\bibitem{AF11}
Alvermann A and Fehske H 2011 {\em J. Comp. Phys.\/} {\bf 230} 5930

\bibitem{PFTV86}
Press W~H, Flannery B~P, Teukolsky S~A and Vetterling W~T 1986 {\em Numerical
  Recipes\/} (Cambridge: Cambridge University Press)

\bibitem{Bo01}
Boyd J~P 2001 {\em Chebyshev and Fourier Spectral Methods\/} (Dover
  Publications)

\bibitem{MQ02}
McLachlan R~I and Quispel G~R~W 2002 {\em Acta Numerica\/}  341

\bibitem{PL86}
Park T~J and Light J~C 1986 {\em J. Chem. Phys.\/} {\bf 85} 5870

\bibitem{HL97}
Hochbruck M and Lubich C 1997 {\em SIAM J. Numer. Anal.\/} {\bf 34} 1911

\bibitem{TK84}
Tal-Ezer H and Kosloff R 1984 {\em J. Chem. Phys.\/} {\bf 81} 3967

\bibitem{HLW06}
Hairer E, Lubich C and Wanner G 2006 {\em Geometric Numerical Integration\/}
  (Berlin: Springer)

\bibitem{Thal06}
Thalhammer M 2006 {\em SIAM J. Numer. Anal.\/} {\bf 44} 851

\bibitem{LGCC09}
De~Liberato S, Gerace D, Carusotto I and Ciuti C 2009 {\em Phys. Rev. A\/} {\bf
  80} 053810

\bibitem{BERB12}
Bastidas V~M, Emary C, Regler B and Brandes T 2012 {\em Phys. Rev. Lett.\/}
  {\bf 108} 043003

\bibitem{Di54}
Dicke R~H 1954 {\em Phys. Rev.\/} {\bf 93} 99

\bibitem{SZ97}
Scully M~O and Zubairy M 1997 {\em Quantum Optics\/} (Cambridge University
  Press)

\bibitem{LEKMSS04}
Littlewood P~B, Eastham P~R, Keeling J~M~J, Marchetti F~M, Simons B~D and
  Szymanska M~H 2004 {\em J. Phys. Condens. Matter\/} {\bf 16} S3597

\bibitem{EL06}
Eastham P~R and Littlewood P~B 2006 {\em Phys. Rev. B\/} {\bf 73} 085306

\bibitem{SVG94}
Shi H, Verechaka G and Griffen A 1994 {\em Phys. Rev. B\/} {\bf 50} 1119

\bibitem{SS10}
Stolz H and Semkat D 2010 {\em Phys. Rev. B\/} {\bf 81} 081302(R)

\bibitem{Sno06}
Snoke D 2006 {\em Nature\/} {\bf 443} 403

\bibitem{HJ08}
Haug H and Jauho A~P 2008 {\em Quantum Kinetics in Transport and Optics of
  Semiconductors\/} (Berlin Heidelberg New-York: Springer)

\bibitem{Sid98}
Sidje R~B 1998 {\em ACM Trans. Math. Softw.\/} {\bf 24} 130

\bibitem{GNZBS09}
Guan X, Noble C, Zatsarinny O, Bartschat K and Schneider B 2009 {\em Computer
  Physics Communications\/} {\bf 180} 2401

\bibitem{AF08}
Alvermann A and Fehske H 2008 {\em Phys. Rev. B\/} {\bf 77} 045125

\bibitem{AF09}
Alvermann A and Fehske H 2009 {\em Phys. Rev. Lett.\/} {\bf 102} 150601

\bibitem{CRHP10}
Chin A~W, Rivas \'{A}, Huelga S~F and Plenio M~B 2010 {\em J. Math.
  Phys.\/} {\bf 51} 092109

\end{thebibliography}
\end{document}